\documentclass[desactivate,twocolumn]{aa}  

\DeclareRobustCommand{\VAN}[3]{#2}
\let\VANthebibliography\thebibliography
\def\thebibliography{\DeclareRobustCommand{\VAN}[3]{##3}\VANthebibliography}

\usepackage[colorlinks=true,
    linkcolor=blue, citecolor=blue, filecolor=blue, urlcolor=blue]{hyperref}
\usepackage{caption}
\usepackage{subcaption}
\usepackage[version=4]{mhchem}
\usepackage{graphicx}	
\usepackage{amsmath}	
\usepackage{amssymb}	
\usepackage{array}
\usepackage{multirow}
\usepackage{bigdelim}
\usepackage[varg]{txfonts}
\usepackage{xcolor}

\definecolor{mygrey}{RGB}{46, 46, 46}
\definecolor{mywhite}{RGB}{214, 214, 214}
\definecolor{myblue}{RGB}{108, 153, 187}
\definecolor{myred}{RGB}{176, 82, 121}

\begin{document}

\title{The High-resolution Accretion Disks of Embedded protoStars (HADES) simulations. I. Impact of protostellar magnetic fields on accretion modes}
\titlerunning{HADES I: Protostar accretion}
\authorrunning{Gaches et al.}

\author{Brandt A. L. Gaches
        \inst{1}\thanks{E-mail: brandt.gaches@chalmers.se}
        \and
        Jonathan C. Tan
        \inst{1}\fnmsep\inst{2}
        \and
        Anna L. Rosen
        \inst{3}\fnmsep\inst{4}
        \and
        Rolf Kuiper
        \inst{5}
        }
\institute{
Department of Space, Earth and Environment, Chalmers University of Technology, Gothenburg SE-412 96, Sweden
\and
Department of Astronomy, University of Virginia, 530 McCormick Road, Charlottesville, VA 22904, USA
\and
Department of Astronomy, San Diego State University, San Diego, CA 92182, USA
\and
Computational Science Research Center, San Diego State University, San Diego, CA 92182, USA
\and
Faculty of Physics, University of Duisburg-Essen, Lotharstraße 1, 47057, Duisburg, Germany
}

\date{Accepted XXX. Received YYY; in original form ZZZ}

\abstract
{How embedded, actively accreting low-mass protostars accrete mass is still greatly debated. Observations are now piecing together the puzzle of embedded protostellar accretion, in particular with new facilities in the near-infrared. However, high-resolution theoretical models are still lacking, with a stark paucity of detailed simulations of these early phases. Here, we present high-resolution nonideal magnetohydrodynamic simulations of a solar mass protostar accreting at rates exceeding 10$^{-6} M_{\odot}$ yr$^{-1}$. We show the results of the accretion flow for four different protostellar magnetic fields, 10 G, 500 G, 1 kG, and 2 kG, combined with a disk magnetic field. For weaker (10 G and 500 G) protostar magnetic fields, accretion occurs via a turbulent boundary layer mode, with disk material impacting the protostar surface at a wide range of latitudes.  In the 500 G model, the presence of a magnetically dominated outflow focuses the accretion toward the equator, slightly enhancing and ordering the accretion. For kilogauss magnetic fields, the disk becomes truncated due to the protostellar dipole and exhibits magnetospheric accretion, with the 2 kG model having accretion bursts induced by the interchange instability. We present bolometric light curves for the models and find that they reproduce observations of Class I protostars from YSOVAR, with high bursts followed by an exponential decay possibly being a signature of instability-driven accretion. Finally, we present the filling fractions of accretion and find that 90\% of the mass is accreted in a surface area fraction of 10-20\%. These simulations will be extended in future work for a broader parameter space, with their high resolution and high temporal spacing able to explore a wide range of interesting protostellar physics.}


\maketitle

\section{Introduction}
Protostars are the bloated embryos of main-sequence stars that are actively accreting material from their natal environments. In the earlier phases of evolution, low-mass protostars are still deeply embedded within their natal cores and surrounded by relatively massive disks, resulting in substantial accretion rates, $\dot{m}_* \geq 10^{-7}M_{\odot}$ yr$^{-1}$. Direct knowledge of how protostars accrete mass typically comes from observations at more evolved stages, such as T-Tauri stars, when the central protostar is exposed, the disks have been largely depleted and the accretion rates drop below $\dot{m}_* \leq 10^{-8}M_{\odot}$ yr$^{-1}$. Since these objects are exposed, it becomes significantly more feasible to observe emission from the accretion zones, primarily in the form of X-rays \citep{Feigelson1999, Feigelson2007}, ultraviolet emission \citep{Manara2023}, and hydrogen recombination emission \citep{Fischer2023}. There are indirect methods to estimate the accretion rate, such as determining the properties of protostellar outflows and inferring the accretion rate via theoretical models \citep[e.g.,][]{Avison2021}. Modern observational facilities, such as the Very Large Telescope Interferometer (VLTI) with the GRAVITY instrument, are now probing the accretion of more embedded systems, tracing evolutionary times when most of the protostar's mass has already been accumulated \cite[e.g.,][]{GravityCollaboration2023a, GravityCollaboration2023b}. During the main accretion phases of protostar formation, the protostar's natal core and disk obscure the accretion zone, complicating efforts to directly probe the inner accretion disks and follow how the material is delivered onto the protostar. 

Multi-physics star formation simulations can capture the earliest phases of star formation from large scales (sub-parsecs for core-scale simulations and 1-10s of parsecs for molecular cloud simulations) \citep[e.g.,][]{Dale2012, Rosen2016, Offner2017, Seifried2017, Cunningham2018, Wurster2019, Rosen2020, Grudic2021, Rosen2022, Suin2024}, but must use sub-grid prescriptions to model the protostar as an unmagnetized sink particle that accretes mass from an assumed accretion region \citep[][]{Offner2009, Bleuler2014}. These simulations are broadly able to reproduce the accretion rates from analytic accretion models \citep[i.e.,][]{Shu1977, Bonnell2001, McKee2003}. However, these parsec-scale simulations cannot resolve the accretion flow directly onto the protostar due to resolution constraints: resolving from molecular cloud scales to the protostellar surface requires spanning 7-8 orders of magnitude across length scales. The combined theoretical and observational challenges mean the underlying physical processes of protostellar accretion have been computationally and observationally unfeasible to probe during the main growth phases.

Several different regimes of protostellar accretion have been proposed. Boundary layer accretion \citep{Kley1996} occurs when the surrounding disk reaches the protostellar surface, directly feeding material onto the protostar. Magnetospheric accretion \citep{Koenigl1991, Matt2005, Rosen2012, Hartmann2016} occurs when the magnetic pressure associated with the protostar's surface magnetic field truncates the disk, balancing the ram pressure of the accretion flow, funneling the gas along magnetic flux tubes from the disk to the protostellar surface. Analytic expressions for the radius at which the disk is truncated typically assume the field consists solely of a dipole, although the magnetic field of evolved protostars is found to be a complex mixture of multipoles \citep{Donati2007, Donati2008, Gregory2008, Carroll2012}. The flow from the truncated disk is still likely dominated by the dipole component, with the higher-order poles perturbing the flow closer to the surface \citep{Lovelace2010}. These filaments impact the protostar and create a shock near the surface, shock-heating up to millions of degrees Kelvin causing the accretion shock to primarily radiate X-rays via thermal Bremsstrahlung cooling \citep{Calvet1998, Hartmann2016}. Accretion can also occur via funnel flows along the outflow cavity wall and through failed winds \citep{Takasao2018} that occur when outflowing gas fails to escape the protostar's gravitational potential. Finally, instability-driven accretion can occur via the development of fluid instabilities such as the magnetic Rayleigh-Taylor instability (also called the interchange instability) at the truncated disk surface \citep[e.g.,][]{Kulkarni2008, Zhu2024}. For actively accreting, embedded protostars, the underlying mechanisms remain largely unknown, motivating the need for high-resolution multidimensional numerical simulations of accretion disks and their host protostars to resolve the accretion flow onto the protostar. 

High-resolution simulations of protostellar accretion have primarily focused on later evolutionary times, in particular the T-Tauri phase when the protostar is exposed, primarily motivated by the availability of observational data \citep{Zanni2009, Romanova2011, Yuan2012, Zanni2013, Blinova2016, Takasao2018, Ireland2021, Romanova2021, Takasao2022, Zhu2024}. In these simulations, the disk mass is typically less than 1\% of the protostar's mass, resulting in a thin, weakly turbulent, viscous disk. The resulting accretion rates range between $\dot{m}_* \approx 10^{-11} - 10^{-8}$ M$_{\odot}$ yr$^{-1}$. Conversely, at the earliest evolutionary times, high-resolution multi-physics simulations have been performed of the collapse of a protostar core to just after its initial formation, only following the first and second collapse phases to stellar densities \citep{Tomida2013, Bate2014, Bhandare2020, Ahmad2024}. There is, thus, a significant gap in the simulations of embedded protostars during the main growth phase, when the protostar is actively accreting the majority of its stellar mass at rates of $\dot{m}_* \gtrsim 10^{-7}M_{\odot}$ yr$^{-1}$, which resolve the underlying accretion flow onto the protostellar surface.

To address this gap, we performed a suite of high-resolution nonideal viscous magnetohydrodynamic simulations (MHD) of accretion onto a solar mass protostar to bridge this important gap. In order to determine the importance of how the protostar's magnetic field affects the accretion flow structure and delivery of material onto the protostellar surface, we varied the protostar's magnetic field across a wide physical range and included nonideal effects. Our simulations include Ohmic dissipation and the Hall effect, but we neglect ambipolar diffusion in this study. While ambipolar diffusion can be important on the larger disk and protostellar core scales \citep{Lesur2014, Commercon2022}, the Hall effect likely dominates in the inner accretion disk \citep{Lesur2021a}. The role of ambipolar diffusion in the inner hot accretion disk around protostars, in particular because the inner regions are more strongly ionized than the rest of the disk and core, has not been well investigated, compared to the larger scales at which the gas is colder and more neutral. 

Our simulations presented herein focus on a 1 M$_{\odot}$ bloated protostar analog with accretion rates ($\dot{m} \geq 10^{-7}$ M$_{\odot}$ yr$^{-1}$) coinciding with recent observational surveys of \citet{Fiorellino2022, Fiorellino2023} of young, accreting protostars. Since theoretical studies have shown that the protostellar magnetic field plays a major role in the underlying accretion mechanism in evolved protostellar objects (see references above), we focus primarily on varying the protostellar magnetic field from 10 G to 2 kG and detailing the resulting accretion modes. Our Paper II, Chowdhury et al. (in prep.), will detail more quantitatively the outflow physics. In Section \ref{sec:methods} we describe the physics included in the simulations and the initial and boundary conditions imposed. In Section \ref{sec:res} we present our results of the simulations and describe the accretion flow onto the protostar and the interplay between the protostar's accretion rate and and bolometric luminosity variability. In Section \ref{sec:disc} we discuss the results of these numerical experiments in the context of currently known protostar physics. Finally, we conclude and summarize our results in Section \ref{sec:conclude}.

\section{Methods}\label{sec:methods}
We performed four simulations amounting to numerical experiments to investigate the role of the protostellar magnetic field on the underlying accretion physics from its surrounding protostellar disk onto the protostellar surface. We used the public {\sc Pluto} MHD simulation code, which includes nonideal MHD terms for Ohmic resistivity and the Hall effect, shear viscosity, and an external gravitational potential. We neglected ambipolar diffusion in this study. As was stated above, we neglect ambipolar diffusion in our simulations because it is likely subdominant to the other nonideal effects for the innermost regions of the protostellar accretion disk. We used the HLL solver with the Lagrange multiplier divergence cleaning method, to enforce $\nabla\cdot\vec{B} = 0$, and the van Leer limiter. Pressure was updated based on the entropy of the gas, which was advected as a passive scalar. The equations were evolved using a second-order Runge-Kutta integration scheme. The equations solved by {\sc Pluto} are
\begin{equation}
    \frac{\partial \rho}{\partial t} + \nabla \cdot ( \rho \vec{v}) = 0,
\end{equation}
\begin{equation}
    \frac{\partial (\rho\vec{v})}{\partial t} + \nabla \cdot \left [ \rho\vec{v}\vec{v} - \vec{B}\vec{B} + \mathbb{I} \left ( p + \frac{B^2}{2}  \right )\right ]^T = -\rho \nabla \Phi + \nabla \cdot \Pi,
\end{equation}
\begin{equation}
    \frac{\partial \vec{B}}{\partial t} + \nabla \times \left (c\vec{E} \right ) = 0,
\end{equation}
\begin{equation}
    \frac{\partial (E_t + \rho \Phi)}{\partial t} + \nabla \cdot \left [ \left ( \frac{\rho \vec{v}^2}{2} + \rho e + p + \rho \Phi \right )\vec{v} + c\vec{E}\times\vec{B} \right ] = \nabla \cdot (\vec{v}\cdot \Pi),
\end{equation}
\begin{equation}
    \frac{\partial (\rho s)}{\partial t} + \nabla \cdot \left ( \rho s \vec{v} \right ) = 0,
\end{equation}
where $\rho$ is the mass density, $\vec{v}$ is the gas velocity, $\vec{B}$ is the magnetic field, $\vec{E}$ is the induced electric field, $p$ is the gas thermal pressure, $s$ is the gas entropy, and $E_t = \rho e + \vec{m}^2/2\rho + \vec{B}^2/2$ is the total energy density. The entropy was evolved as a passive scalar, since source parabolic terms are always added to the total energy. We treated the fluid as an ideal gas for the closure relation between $\rho$, $e$, and $s$, where we took $\gamma = 5/3$. The gas pressure was recovered through the entropy; that is, $s = \rho/p^\gamma$. $\Phi$ is the external gravitational potential originating from the central protostar, described in Section \ref{sec:init}. 

The induced electric field, $\vec{E}$, is
\begin{equation}
    c\vec{E} = -(\vec{v} + \vec{v}_H) \times \vec{B} + \frac{\eta}{c}\cdot\vec{J},
\end{equation}
where $\vec{v}_H = -\vec{J}/en_e$, $e$ is the elementary charge, $n_e$ is the electron number density, $\vec{J} = c\nabla \times B$ is the induced current, and $\eta$ is the magnetic resistivity. Our implementation of the magnetic resistivity, only including Ohmic dissipation, and our prescription for the ionization fraction in described in Section \ref{sec:visc}. The term $\vec{v}_H$ introduces the Hall effect into the nonideal MHD equations. We included shear viscosity, $\nu_s$, which was coupled to the above equations with the viscous stress tensor
\begin{equation}
    \Pi = \eta_s \left [ \nabla \vec{v} + (\nabla \vec{v})^T \right ] - \frac{2}{3}\eta_s \left ( \nabla \cdot \vec{v} \right ) \mathbb{I}.
\end{equation}

We solved these equations in axisymmetric spherical coordinates (r, $\theta$), including all three components of vectors (2.5D). We used 512 radial logarithmically spaced cells from the protostellar surface out to 1 AU. In the $\theta$ direction, we used 512 cells between $[10^{-2}, \pi-10^{-2}]$ radians to avoid the poles at $\theta = 0, \pi$. This leads to $\Delta r \approx 1\times10^{-4}$ AU at the protostar surface and $\Delta r \approx 8\times10^{-3}$ AU at the 1 AU boundary. The high resolution near the protostellar surface is essential to capture the accretion flow structure and turbulence, which is lost at lower resolution. The inner bloated protostar surface is $R_* = 3 {\,\rm R_{\odot}} = 0.01395$ AU, motivated by models accreting bloating solar mass protostars \citep[][]{Geroux2016}),

\subsection{Initial conditions}\label{sec:init}
We followed the convention that $(r,~\theta,~\phi)$ denotes the spherical coordinates, and the two-dimensional cylindrical coordinates are $(R = r\sin\theta,~z = r\cos\theta)$. The initial disk surface density profile is the often-used truncated-power law \citep[e.g.,][]{Lynden-Bell1974, Bethune2017, Zhu2024, Rea2024}. We assumed an initial disk surface density given by a purely viscous disk, 
\begin{equation}
    \Sigma(R) = \Sigma_0 \left ( \frac{R}{R_0}\right )^{-1} e^{-R/R_{\rm max}},
\end{equation}
where $\Sigma$ is the gas surface density, $\Sigma_0$ is the normalizing factor, described below, $R$ is the cylindrical radius, and $R_0$ is the normalization radius, taken to be $R_0 = 10$ AU. The initial disk was assumed to be in hydrostatic balance, such that the gas density is
\begin{equation}
    \rho = \frac{\Sigma(R)}{\sqrt{2\pi}h} \exp{\left (-\frac{z^2}{2h^2}\right )},
\end{equation}
where $h = c_s/\Omega_K$ is the scale height of the disk and $\Omega_K = \sqrt{Gm_*/R^3}$ is the Keplerian rotation rate. The disk extends out to $R_{\rm max} = 40$ AU, consistent with survey results of protostars in the Orion molecular cloud similar to our model \citep{Tobin2020, Sheehan2022}. Our initial disk surface density profile is marginally steeper than the average of the modeled disks in \citet{Sheehan2022} but is still within the range of their survey. The tangential velocity is the Keplerian speed,
\begin{equation}
    v_{\phi} = \sqrt{\frac{Gm_*}{R}}.
\end{equation}
There is a prescribed initial radial velocity as a function of the target accretion rate, $\dot{m}_*$, assuming the accretion onto the surface is the same as through the disk, $\dot{m}_* = \dot{m}_{\rm disk}$. The radial velocity is set by mass conservation through the disk,
\begin{equation}\label{eq:vr_init}
    v_r = -\frac{\dot{m}_{\rm disk}}{2\pi\Sigma(R)R}.
\end{equation}

Finally, there is an initial core described by a purely infalling envelope \citep{Ulrich1976}, with a density profile,
\begin{equation}
    \rho_{\rm env} = \rho_{\rm 0, env} \left ( \frac{r}{50 \,{\rm AU}} \right )^{-1.5}.
\end{equation}
While the initial core is disrupted in the simulation by outflowing gas, it can remain in the outer domain and helps balance the pressure of the disk in the initial condition. 

The values of $\Sigma_0$ and $\rho_{\rm 0, env}$ are determined to get a total mass ratio between the protostar and its natal disk and core, such that $m_{\rm disk}/m_* = 0.175$ for the whole disk (out to $R_{\rm max} = 40$ AU) and $m_{\rm core}/m_* = 0.1$ (out to 0.1 pc). 

The initial temperature was assumed to vary with cylindrical radius as
\begin{equation}
    T(r) = T_0 \left (  \frac{R}{R_0} \right )^{-0.5},
\end{equation}
where $T_0 = 200\:$K. This temperature is warmer than the recommended relation from \citet{Sheehan2022} using only the intrinsic protostellar luminosity, but it is consistent if the accretion luminosity corresponding to the targeted accretion rate is added as an additional luminosity component. The temperature radial scaling exponent can take a range of values \citep{Kenyon1993, Chiang1997}, and a value of $-0.5$ is consistent with radiative transfer modeling of embedded protostellar disks \citet{Tobin2020}.

Both the protostar and the disk are magnetized. We defined an initial disk magnetic field similar to \citet{Zanni2009}, such that the magnetic field is vertical in the outer disk and transitions to an hourglass morphology in the inner disk. The magnetic field was initialized by the vector potential, $\vec{A}$, where $\vec{B} = \nabla \times \vec{A}$. Given the axisymmetry, we only initialized the $\phi$-component:
\begin{equation}
A_{\phi} =  \frac{2B_1 R_1}{3 - a} r^{-(a - 1)/2}, 
\end{equation}
where $B_1 = \sqrt{2p_1/\beta_1}$, $p_1$ is the pressure at 1 AU, $R_1 = 1$ AU, and $\beta_1 = p_1/p_{\rm 1, mag}$ is plasma beta, the ratio of thermal pressure to magnetic pressure, calculated at 1 AU. We used $a = 1$ as the simplest initial case. This produces an initial poloidal magnetic field which has an hourglass morphology in the inner regions and transitions to a vertical field by $R\approx 0.5$ AU. We used $\beta = 10^5$, which leads to a magnetic field at the pole of $\approx$ 200 G. The protostellar magnetic field is included as a background dipole magnetic field, derived by the magnetic vector potential 
\begin{equation}
    A_{\phi,*} = \frac{B_* R_*^3}{r^2},
\end{equation}
where $B_*$ is the surface magnetic field strength. We explored four different protostellar magnetic fields, with $B_* = $10, 500, 1000, and 2000 G fields, encompassing the range of observationally inferred surface magnetic fields in the recent survey of low-mass protostars from \citet[][]{Flores2024}. Evolved protostars, in particular T-Tauri stars, have been found to have complex magnetic fields \citep{Donati2007, Donati2008, Gregory2008, Carroll2012}, but there are no constraints on this for embedded protostars. Therefore, we assume a pure dipole, typical of many studies \citep[e.g.,][]{Rosen2012, Zanni2013, Ireland2021, Takasao2022}, but we note that the disk magnetic field will perturb the field around the star, regardless, away from a pure dipole.

We included the protostar's gravitational potential following
\begin{equation}
    \Phi(r) = -\frac{Gm_*}{r},
\end{equation}
such that the gravitational acceleration due to the protostar is $\vec{a}_{\rm grav} = -\nabla\Phi$.  We neglected the disk self-gravity since in this inner 1 AU, over these short timescales, the gravitational potential will be dominated by the central protostar.

The protostar, which is inside the inner boundary, was assumed to be rotating at some fraction, $f_{\rm bu}$, of the breakup speed,
\begin{equation}
\Omega_* = f_{\rm bu} \sqrt{\frac{Gm_*}{R_*^3}},
\end{equation}
where we took $f_{\rm bu} = 1/3$ as a nominal median value between slow- and fast-rotators \citep{White2004, Herbst2007, Rosen2012}. The value of $f_{\rm bu} = 1/3$ gives a rotation rate of $4.026\times 10^{-5}$ s$^{-1}$, or a rotation period of $T_{\rm rot} = 2\pi/\Omega_* \approx 1.56\times 10^5$ seconds. The actual rotation speeds of protostars are largely unknown and will be explored in future simulation suites, in particular for highly magnetized protostars.

\subsection{Viscosity and resistivity}\label{sec:visc}
The shear viscosity was included through an $\alpha$-viscosity parametrization \citep{Shakura1973}, 
\begin{equation}
    \eta_s = \alpha_v c_s h,
\end{equation}
where $c_s$ is the sound speed and $h$ is the scale height, computed by using the local sound speed and Keplerian rotation rate. We set $\alpha_v = 0.01$, which for our initial setup would correspond to a floor accretion rate of $\dot{m} \approx 10^{-7}$ M$_{\odot}$ yr$^{-1}$. This effectively acts as a viscosity minimum, to ensure a minimal degree of accretion and angular momentum loss through the disk. However, at our simulation resolutions, it is possible to resolve turbulence within the inner disk which will act as an additional source of angular momentum loss.

For the nonideal magnetic resistivity, we included Ohmic dissipation following \citet{Balbus2001, Lesur2014},
\begin{equation}
    \eta_{\rm O} = \frac{234}{x_e} \sqrt T {\: \rm cm^2 \, s^{-1}},
\end{equation}
where $x_e$ is the ionization fraction. We computed the ionization fraction assuming two components: thermal ionization, which will be dominated by low ionization potential alkalis Na$^+$ and K$^+$ \citep{Balbus2000}, and X-ray ionization. The electron fraction due to the thermal ionization of Potassium is \citep{Balbus2000, Fromang2002},
\begin{multline}
    x_e^{\rm th}  \approx 6.47\times 10^{-13} \left (\frac{x_K}{10^{-7}} \right )  \left (\frac{T}{10^3} \right )^{0.75} \times \\
     \left ( \frac{2.4 \times 10^{15} {\: \rm cm^{-3}}}{n} \right ) \left (\frac{e^{-(25188 {\:\rm K})/T}}{1.15\times10^{-11}} \right ),
\end{multline}
where $x_K$ is the potassium abundance, taken to be $x_K = 10^{-7}$ \citep{Fromang2002}, and $n$ is the gas hydrogen-nuclei number density. 

For the X-ray ionization, we broke the domain into two components using the initial disk density profile. For regions outside of the disk, defined by $z/h > 4$, we assumed the X-ray flux is optically thin, 
\begin{equation}
    F_X(r) = \frac{L_X}{4\pi r^2},
\end{equation} 
where $L_X$ is an assumed X-ray luminosity, nominally taken to be $L_X = 10^{30}$ erg s$^{-1}$, consistent with a solar mass protostar \citep{Getman2022}. Inside the disk, we attenuated the X-ray emission vertically through the disk using a column density computed by the initial disk profile,
\begin{equation}
    F_X(r) = F_X(R, z) = \frac{L_X}{4\pi R^2} e^{-\sigma_X \int_z^{z_{\rm surf}} n(R,z') dz'},
\end{equation}
where $n(R,z)$ is the number density of the initial disk density distribution, and $\sigma_X \approx 10^{-22}$ cm$^{-2}$ is the X-ray photo-absorption cross section around 1 keV for gas with dominantly neutral hydrogen \citep{Gaches2023} (does not vary much until the gas temperature greatly exceeds $10^4$ K). Since the initial disk is assumed to be in hydrostatic equilibrium, the integral in the exponential has an analytic solution. The X-ray ionization rate, $\zeta$, was computed assuming that it is dominated by secondary ionization, determined by the energy deposition (see below) and the mean energy per ion pair \citep{Dalgarno1999, Meijerink2005}.

The ionization fraction due to X-ray irradiation is then
\begin{equation}
    x_e^{\zeta} = \sqrt{\frac{\zeta}{\beta n}},
\end{equation}
where $\beta$ is the recombination rate for molecular ions \citep{Fromang2002}. By construction, $\zeta$ is constant throughout the disk, but $x_e^{\zeta}$ will vary due to the changing gas density. The total ionization fraction is $x_e = x_e^{\rm th} + x_e^{\zeta}$. The electron density is $n_e = x_e n$. Including a full self-consistent ionization solver is not within the scope of this work, which focuses on the impact of the protostellar magnetic field on the accretion flow. 

We only included the viscosity and resistivity where $r \ge r_{\rm corot}$, where $r_{\rm corot} = \left (\frac{GM_*}{\Omega_*^2} \right )^{1/3}$ is the corotation radius since inside this radius the gas is expected to be ionized enough for the gas to evolve more similarly to ideal MHD. We tapered the resistivity and viscosity to zero from the inner disk boundary to $r_{\rm corot} = 0.029$ AU (for our assumed $f_{\rm bu}$) to avoid a discontinuity in the viscosity and resistivity.

\subsection{Heating and cooling}
We included both atomic line and continuum cooling for gas between $10^4 {\,\rm K} < T < 10^9 {\,\rm K}$ and X-ray gas heating from the central protostar. The line cooling was included as a tabulated cooling rate calculated using the {\sc ChiantiPy} \citep{Dere2013, Dere2019} package for solar abundances and collisional ionization equilibrium. Since we did not include chemistry, we neglected molecular line cooling. Further, we did not include dust cooling, and as such the simulations may be warmer, especially in the disk where the gas temperature can drop below the dust sublimation temperature. X-ray gas heating was included using a simple prescription. The heating term is
\begin{equation}
    \varepsilon_X = n_H H_X \,\, {\rm erg \,s^{-1} \,cm^{-3}},
\end{equation}
where $H_X \approx F_X \sigma_X$ is the X-ray energy deposition rate and $F_X$ is the X-ray flux computed in the manner described above. X-ray heating is necessary to ensure a heated envelope above the protostar and a smoother thermal profile away from the protostar. In the nonideal, viscous MHD equations, there are also heating contributions due to the dissipation of the magnetic field and shear viscosity.

\subsection{Boundary conditions}
The inner boundary must be treated carefully to avoid inducing artifacts in the magnetic field from the rotating, magnetized protostar surface \citep{Zanni2009}. We assumed that the protostar is a perfect conductor, such that $\vec{E} = \vec{B} \times ( \vec{u} - \vec{\Omega_*} \times \vec{R}) = 0$. The protostar was set to be rotating aligned with the $\theta = 0$ pole. For infalling material, we forced the condition that $E_\phi = 0$ by setting the poloidal velocity components according to the poloidal magnetic field. For outflowing material, we set the invariant $k = 4\pi \rho u_p/B_p$ to be constant along the radial boundary. The $B_r$ and $B_\theta$ components used an outflow boundary condition. To ensure we did not induce an electron field inside the magnetic field, we set the toroidal magnetic field by assuming $RB_\phi$ is conserved across the boundary and that $u_\phi = R\Omega_* + u_p B_\phi/B_p$. These criteria ensured that the gas falling onto the protostellar sink surface is absorbed perfectly through the boundary. At the outer radial boundary, we assumed an outflow-but-no-inflow condition. We used axisymmetric boundary conditions in the polar direction. 

\section{Results}\label{sec:res}
The high spatial resolution, from $\approx 10^{-4} - 10^{-2}$ AU, and high-temporal output cadence, $\approx$30 minutes, provides a unique view into the role the protostellar magnetic field plays in the underlying accretion physics. The simulations are evolved for 60 days due to time step constraints caused by the highest wave speeds tending to be in the regions with the highest resolution and to avoid any noticeable disk depletion caused by the lack of inflow at the outer boundary. For the analysis presented below, we ignore the first 11 days of evolution to remove the impact of transients from the relaxation of the initial conditions. This results in 49 days, or approximately 27 protostellar rotation periods, defined by $T_{\rm rot} = 2\pi/\Omega_*$, of evolution after the transient. In the following, we discuss the morphology of the accretion disk, the differences in the accretion variability depending on the strength of the protostellar magnetic field, and examine the accretion flow onto the star and determine its impact location and covering fraction on the protostellar surface.

\subsection{The inner 1 AU: Impact of protostellar magnetic field}
\begin{figure*}
    \centering
    \begin{tabular}{cc}
        \includegraphics[width=0.33\textwidth]{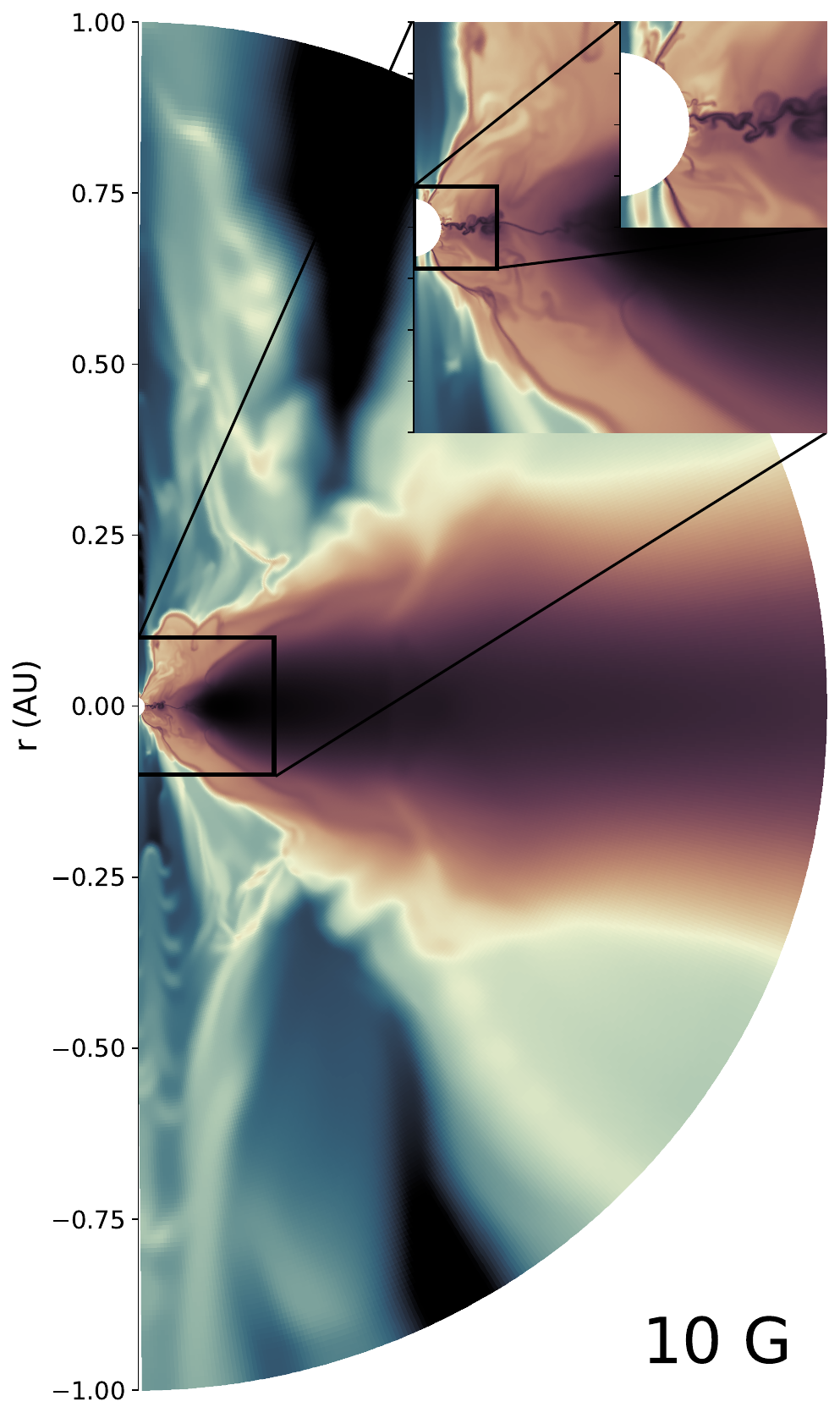} &
        \includegraphics[width=0.33\textwidth]{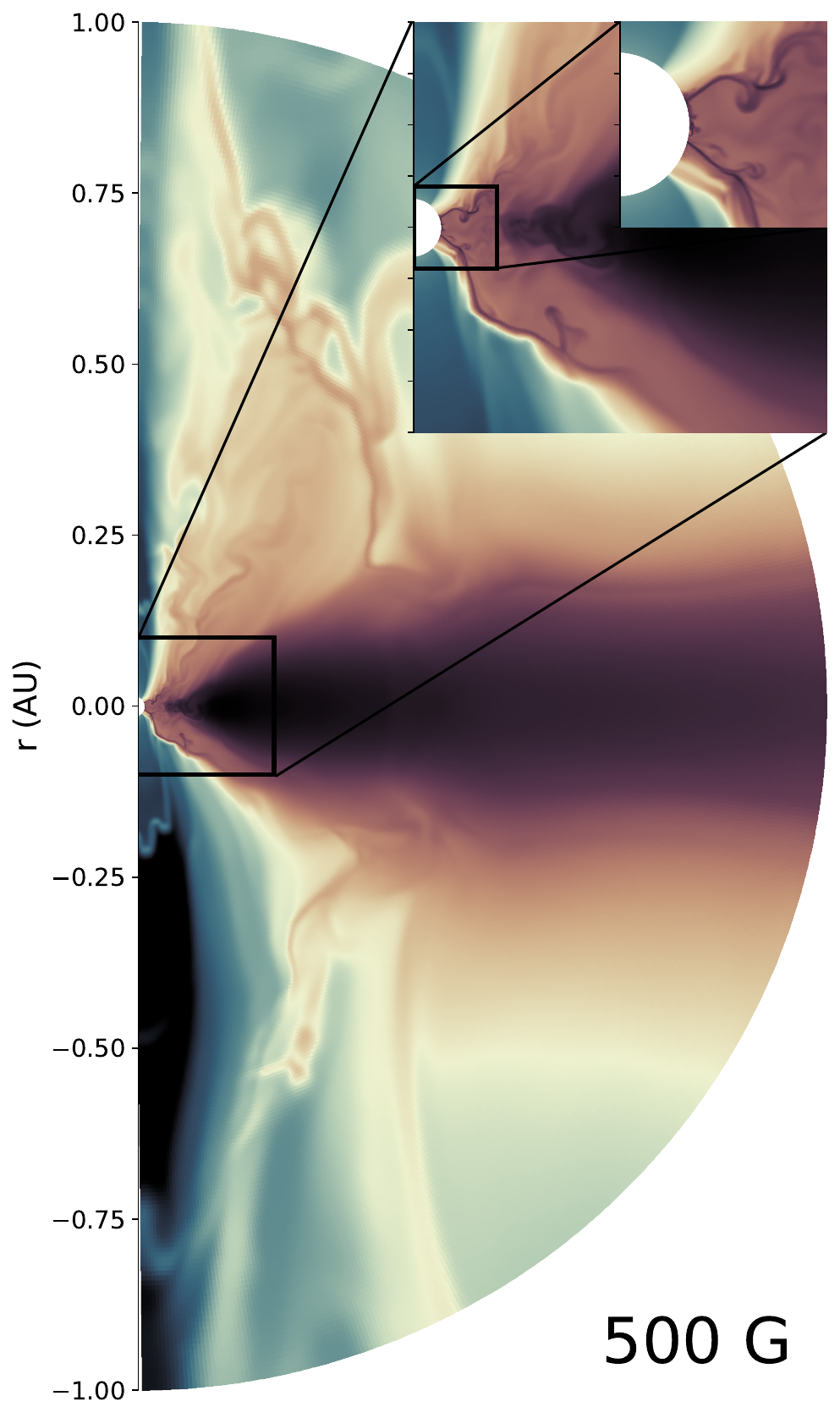} \\
        \includegraphics[width=0.33\textwidth]{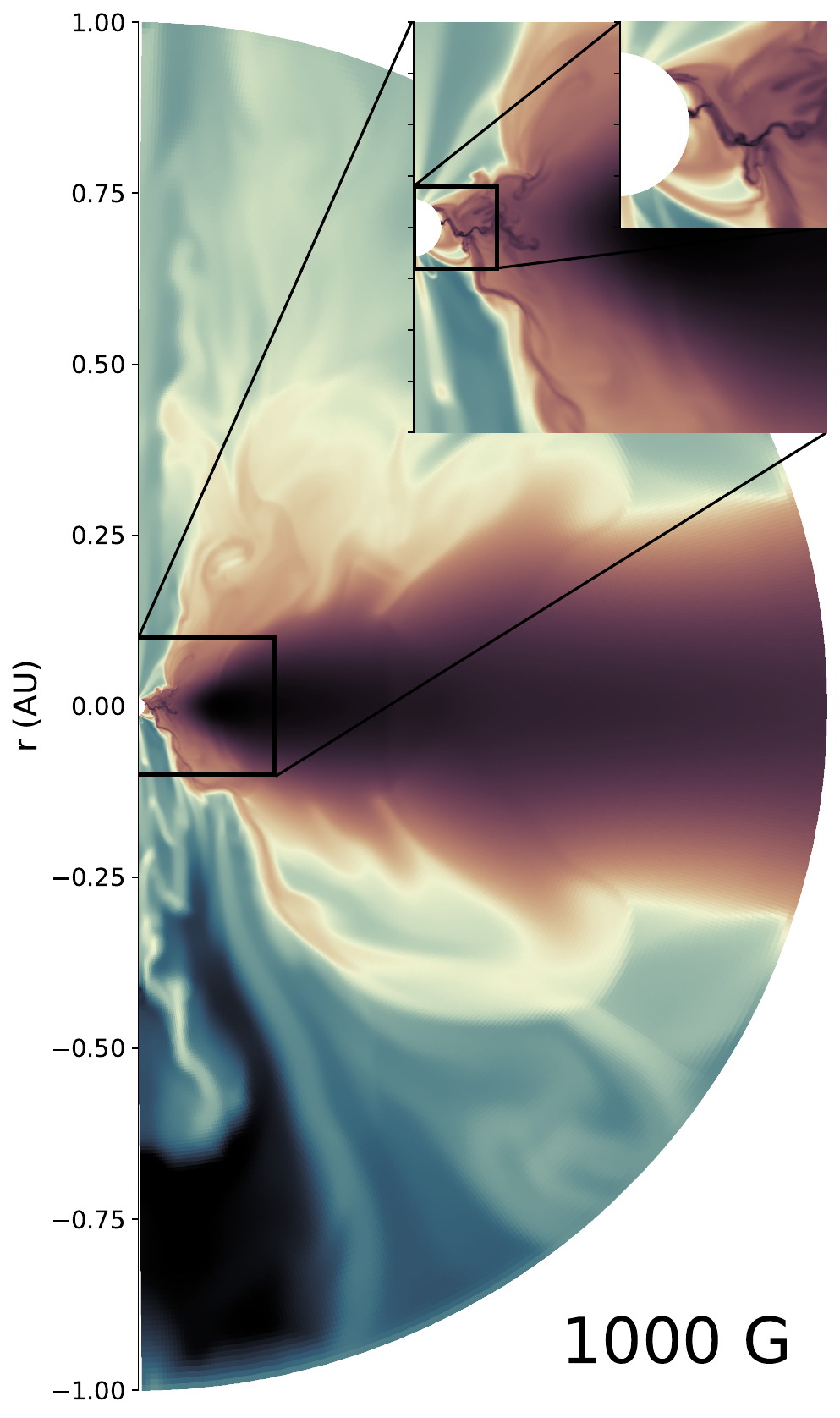} &
        \includegraphics[width=0.33\textwidth]{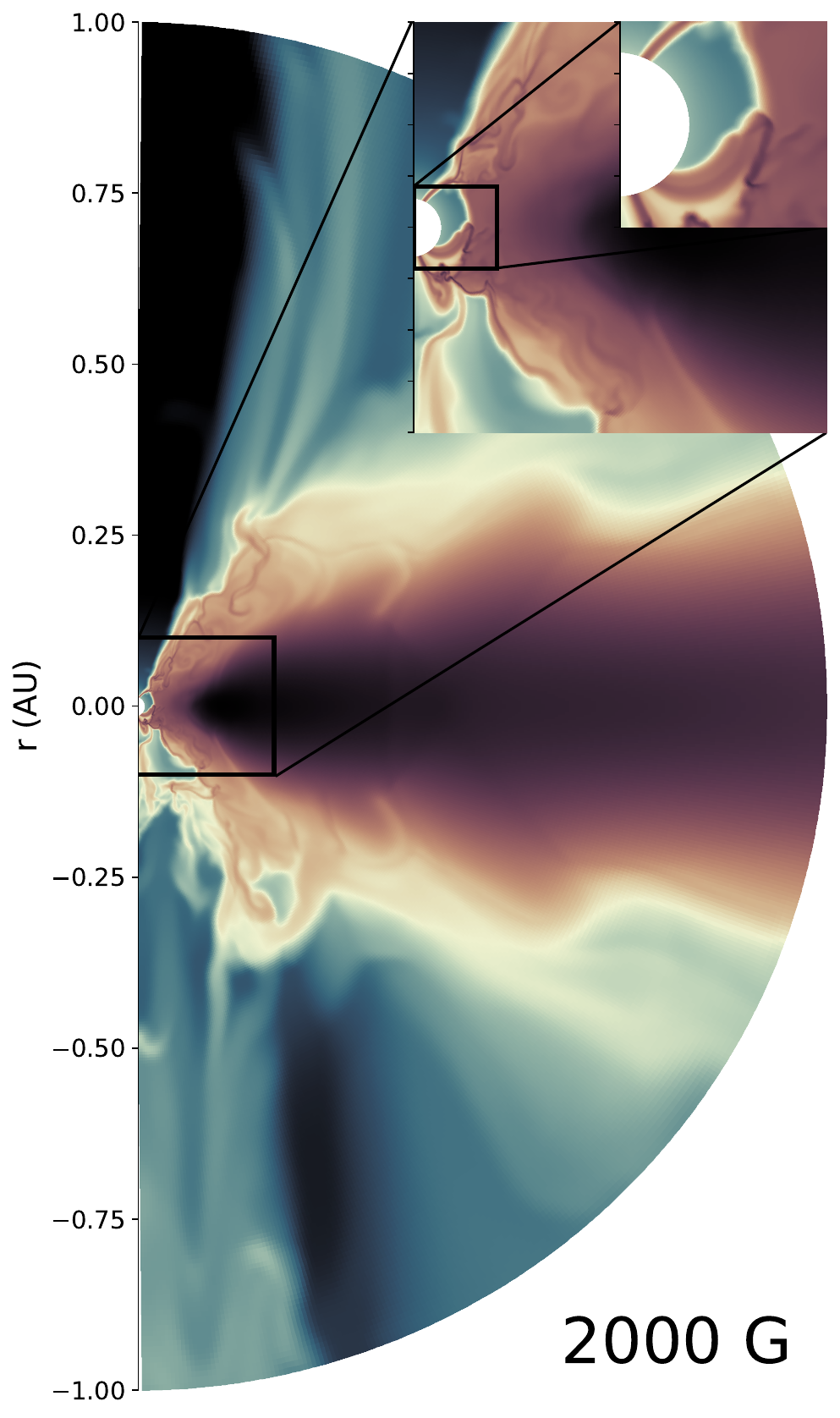}
    \end{tabular}
    \caption{\label{fig:zoom} Density distribution for the 10 Gauss (top left), 500 Gauss (top right), 1000 Gauss (bottom left), and 2000 Gauss (bottom right) protostar, with zoom in varying the box size from 2 AU to 0.04 AU.}
\end{figure*}

\begin{figure}
    \centering
    \includegraphics[width=\columnwidth]{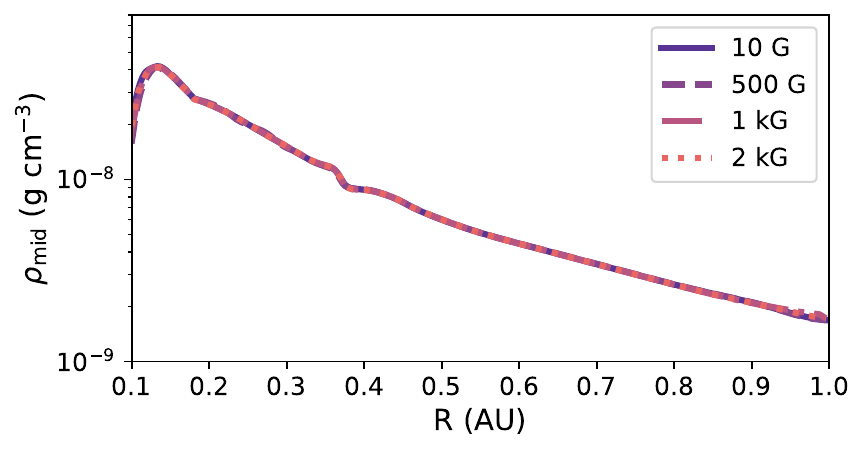}
    \caption{\label{fig:midplaneDens}Gas density, $\rho_{\rm mid}$, versus cylindrical radius, $R$, averaged 22$^{\circ}$ around the midplane, at the final simulation snapshot for the four different protostar magnetic field models, demonstrating that the out disk midplanes are not impacted by the protostellar magnetic field.}
\end{figure}

Figure \ref{fig:zoom} shows the density distribution of the gas from 1 AU scales down to the protostellar surface, with the insets showing progressively higher zoom-ins. The density profiles of the four disks are morphologically similar at the outer radii, as is shown in Figure \ref{fig:midplaneDens}, which displays the density distribution averaged 22$^{\circ}$ around the midplane. This suggests that the disk's bulk is likely influenced more by the disk magnetic field and viscosity effects than by the protostellar magnetic fields. Overall, the disks appear to be lifted, which is expected due to the magnetic buoyancy pulling gas from the disk out as a wind. The upper disk layers and outflow cavities however show significant differences between each other. While all simulations feature outflows, the nature of these appears very different. For the low magnetized protostar, the outflow appears very turbulent, dominated by gas which is thrown off the protostar near the surface. The more magnetized protostars exhibit outflows that are more collimated and ordered, with filaments and knots of gas from magnetospheric ejections near the protostellar surface. Moreover, we also find one-sided outflows, which are caused by breaking the symmetry of the protostellar dipole magnetic field due to its interaction with the disk magnetic field. Such one-sided outflows have been observed in some nearby protostars \citep[e.g.,][]{Federman2024} and simulations of T-Tauri-like protostars with complex fields \citep{Lovelace2010}. 

As we zoom in, the figure shows substantial differences begin to appear between the different protostar magnetic field simulations. The 10 G protostar exhibits a highly turbulent flow within 0.1 AU, with the inflow covering the vast fraction of the protostellar surface. The very weak protostellar dipole means there is substantially less magnetic pressure to counter the ram pressure of the infalling gas. The 500 G, 1 kG, and 2 kG protostars exhibit a more ordered inner 0.1 AU, with a clear differentiation between the magnetically driven bipolar outflow cavity and the accreting gas. For the 1 kG and 2 kG protostar fields, the disk is truncated due to the protostar's magnetic pressure causing gas to flow along the protostar's magnetic field lines where it is funneled to the star in agreement with previous work \citep{Koenigl1991}. These most inner regions can traced by hydrogen recombination emission, particularly in the NIR, hinting that probing these regions with high-resolution spectroscopy could provide a direct test of the accretion physics, as is done for more evolved objects \citep[e.g.,][]{GravityCollaboration2023a, GravityCollaboration2023b}.

\subsection{Location of accretion on the protostellar surface}
Where accretion occurs is of substantial interest in star formation physics. Star formation simulations utilize sink particles with sub-grid protostellar evolution models \citep[e.g.,][]{Offner2009} that predict the quantity of gas accreted, but not how and where. Sub-grid models often utilize ad-hoc prescriptions for the accretion luminosity and launched outflows \citep[e.g.,][]{Cunningham2011, Cunningham2018}. The physical mechanism, covering fraction, and location of the accretion flow onto the surface are also important assumptions utilized in radiation transfer models of protostar accretion \citep[see][and references therein]{Hartmann2016}. 

\begin{figure*}
    \centering
    \includegraphics[width=0.985\textwidth]{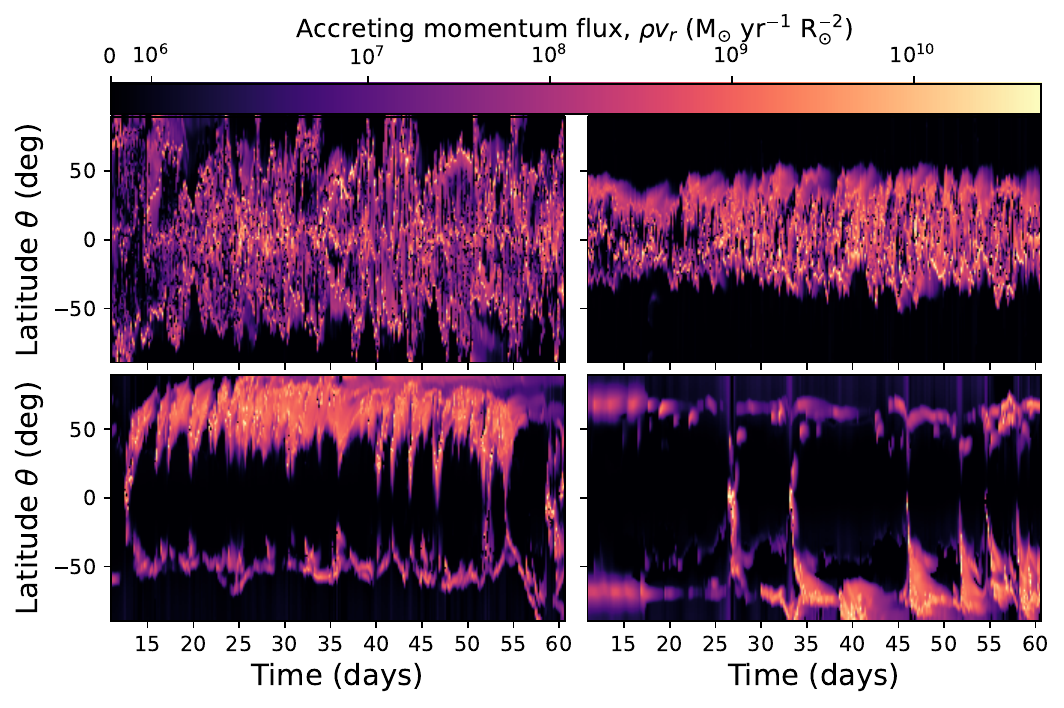}
    \caption{\label{fig:accAreaMap} Accretion rates per unit area across the protostellar surface as a function of time for the four difference cases.}
\end{figure*}

Under the canonical boundary layer and magnetospheric accretion models, one would find that the accretion flow is either dominantly at the equator, or in high latitude accretion columns, respectively. Figure \ref{fig:accAreaMap} summarizes where accretion occurs onto the protostar, via the accretion momentum density (or accretion rate per area, $\rho v_r$). As the protostellar magnetic field increases, the accretion behavior changes in a non-monotonic way. For low protostellar magnetic fields (10 G), the gas falls across the entire surface in a highly chaotic manner, dominated by a filament impacting the equator. At moderate protostellar magnetic fields (500 G), the disk still impacts the protostellar surface directly, with the bipolar outflow providing pressure inhibiting gas flow to high latitudes. Finally, at high protostellar magnetic fields (1 kG and 2 kG), the flow becomes funneled through accretion columns along magnetic field lines, with one accretion column dominating.

\begin{figure*}
    \centering
    \includegraphics[width=0.2\textwidth]{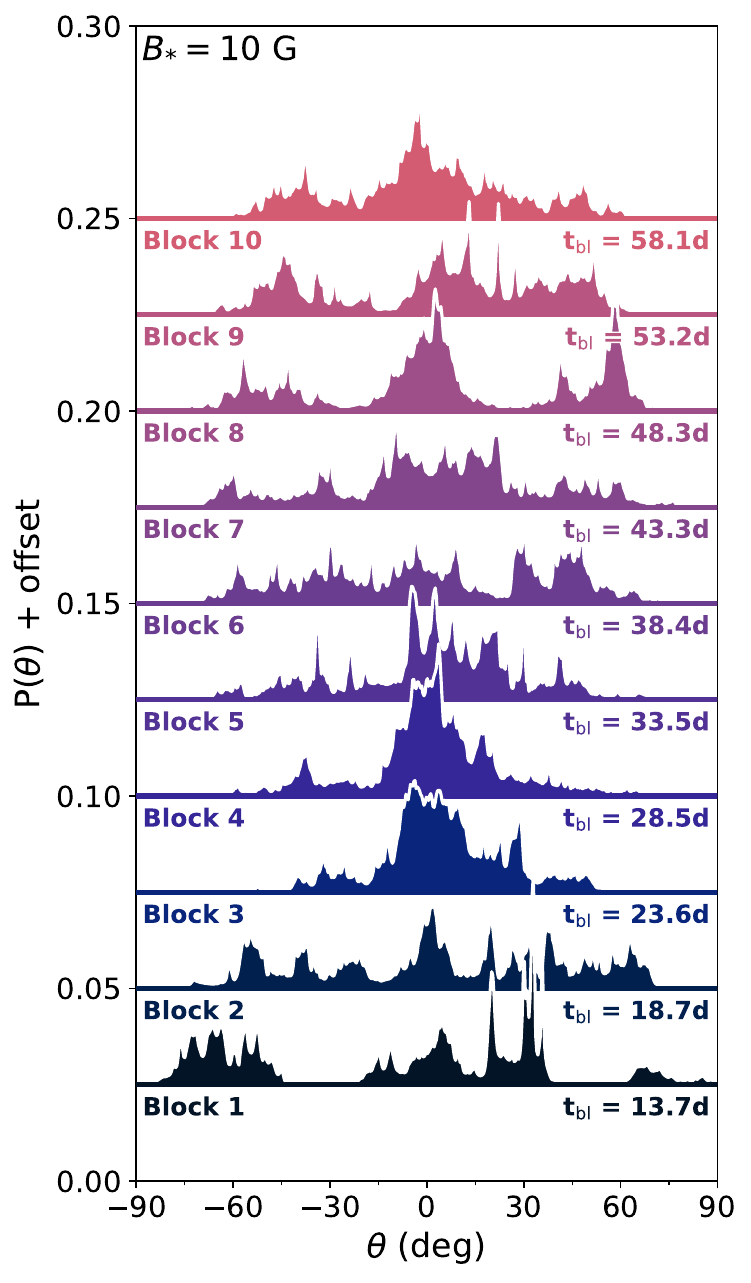}
    \includegraphics[width=0.2\textwidth]{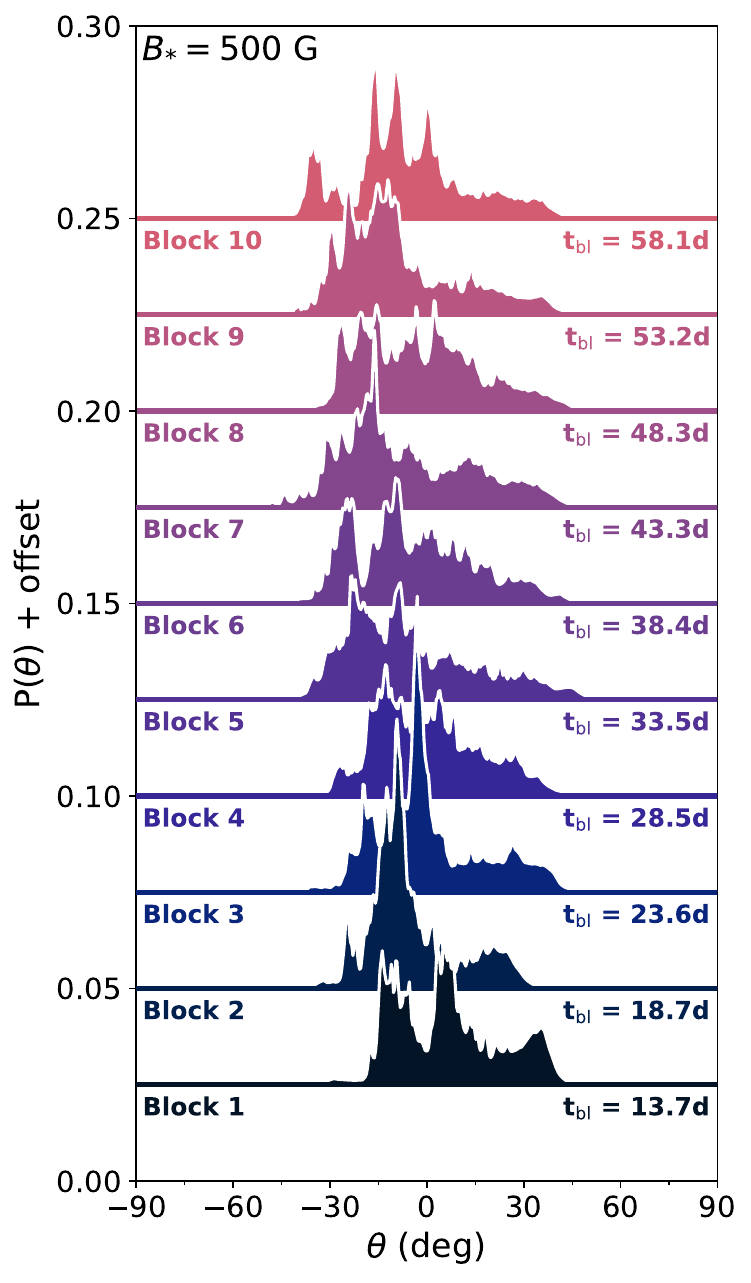}
    \includegraphics[width=0.2\textwidth]{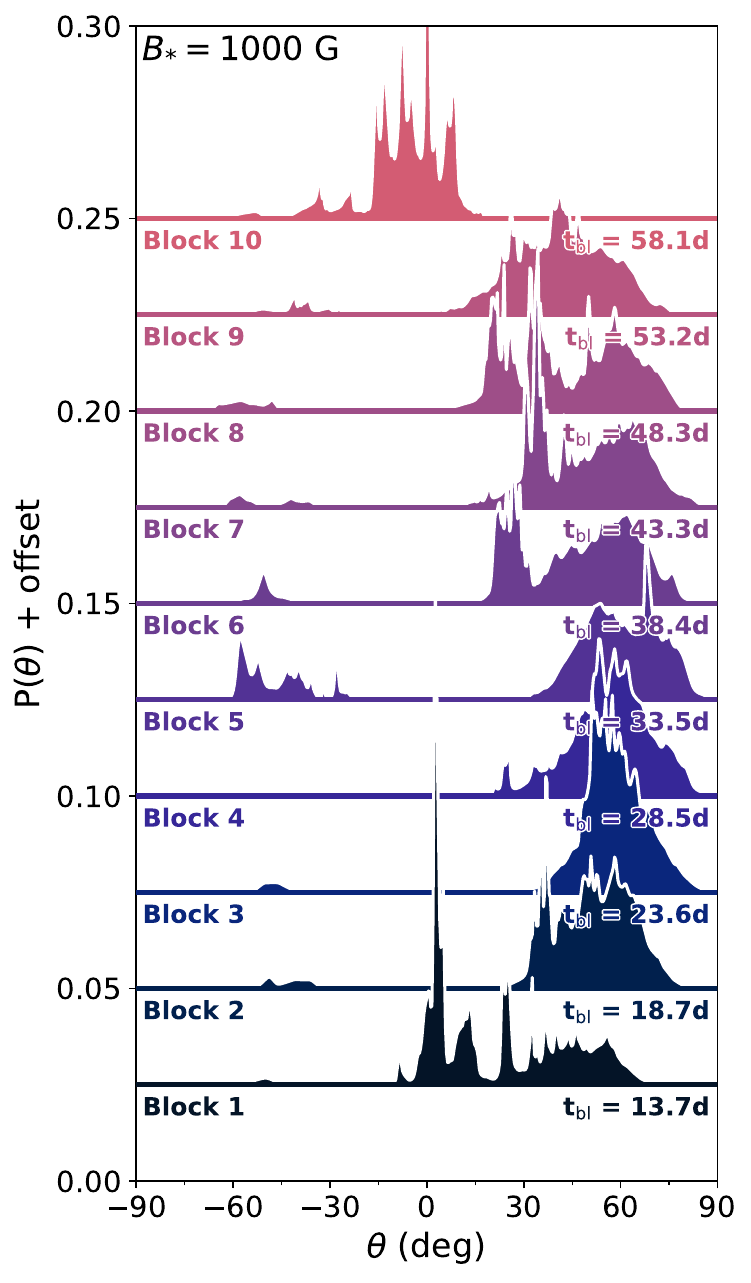}
    \includegraphics[width=0.2\textwidth]{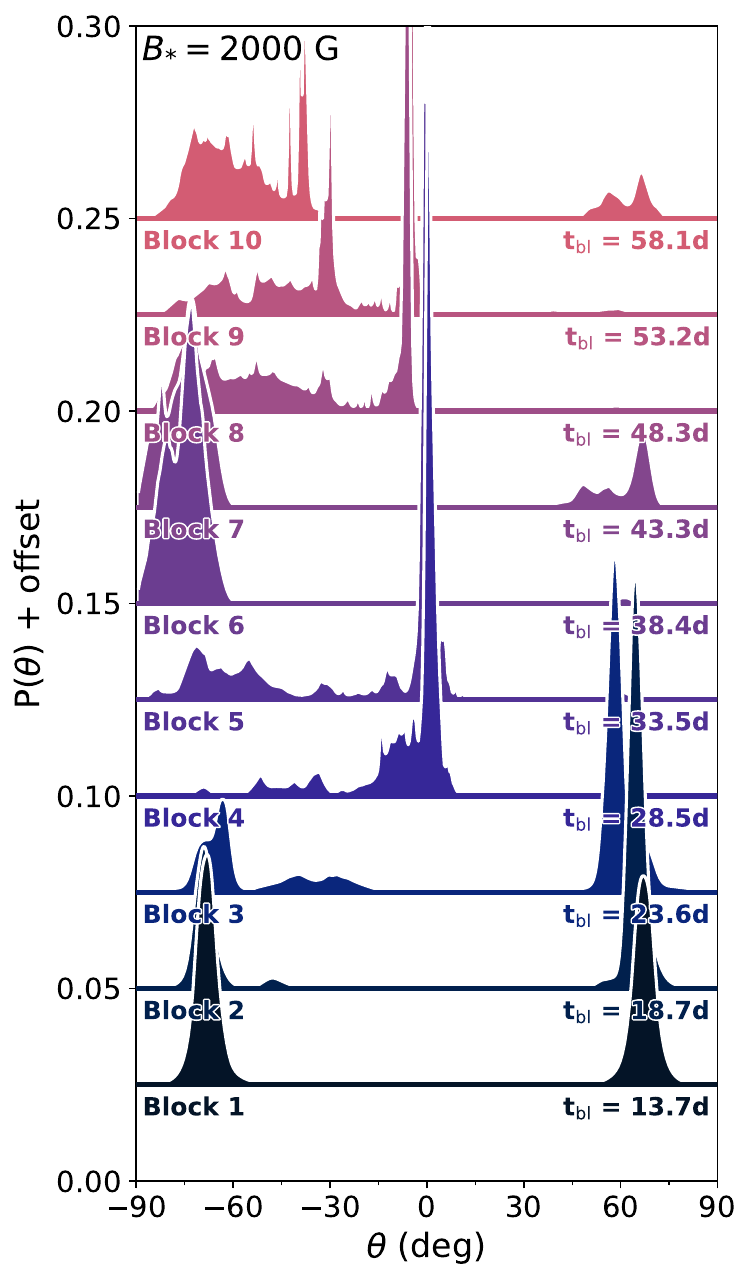}
    \caption{\label{fig:angAcc} Probability distributions of the angles (in degrees) at which mass is accreted onto the central protostar over $\approx$ 5-day blocks for the 10 G (left) 500 G (middle left) and 1000 Gauss (middle right) and 2000 Gauss (right) protostars. Annotated is the block number (also denoted by color), and the time in the center of the block in days.}
\end{figure*}

Figure \ref{fig:angAcc} shows the evolution of time-averaged angular probability distributions of accretion onto the protostar for the different protostellar magnetic fields, shown in different 5-day blocks. The probability is averaging the angular distribution for each snapshot, $i$, in the block, $P_i(\theta) \approx \dot{m}(\theta)/\dot{m}$, where $\dot{m}(\theta)$ is the accretion rate across the surface. The block's central time is annotated. Since the distributions are all normalized to the total gas accretion in each block, we can directly compare the differences between the different magnetic fields and their accretion distribution evolution.

Even from early times, there are substantial systematic differences between the angular accretion distributions between the different magnetic fields. The 10 G protostar field shows a distribution that varies substantially between the time blocks, with periods where the bulk of the gas is accreted at the equator and other blocks where the gas is accreted more evenly distributed across the surface. The higher latitude accretion peaks in the 10 G model are from failed winds falling back along the disk surface. For the 500 G protostar field, the accretion occurs almost entirely as an equatorial flow. The accretion flow is constrained by the strong bipolar outflow, resulting in almost all the accretion occurring within $\pm 30^{\circ}$. At higher protostellar magnetic fields (1 kG and 2 kG) the disk becomes truncated, with an accretion column impacting the protostar broadly at higher latitudes, although in different hemispheres. Lower latitude flows and equatorial accretion occur in the 1 kG and 2 kG simulations during accretion bursts from matter streaming directly from the truncation radius to the protostar (see Section \ref{sec:burst}).

\subsection{Accretion variability and morphology dependence on $B_*$}
\begin{figure}
    \centering
    \includegraphics[width=0.48\textwidth]{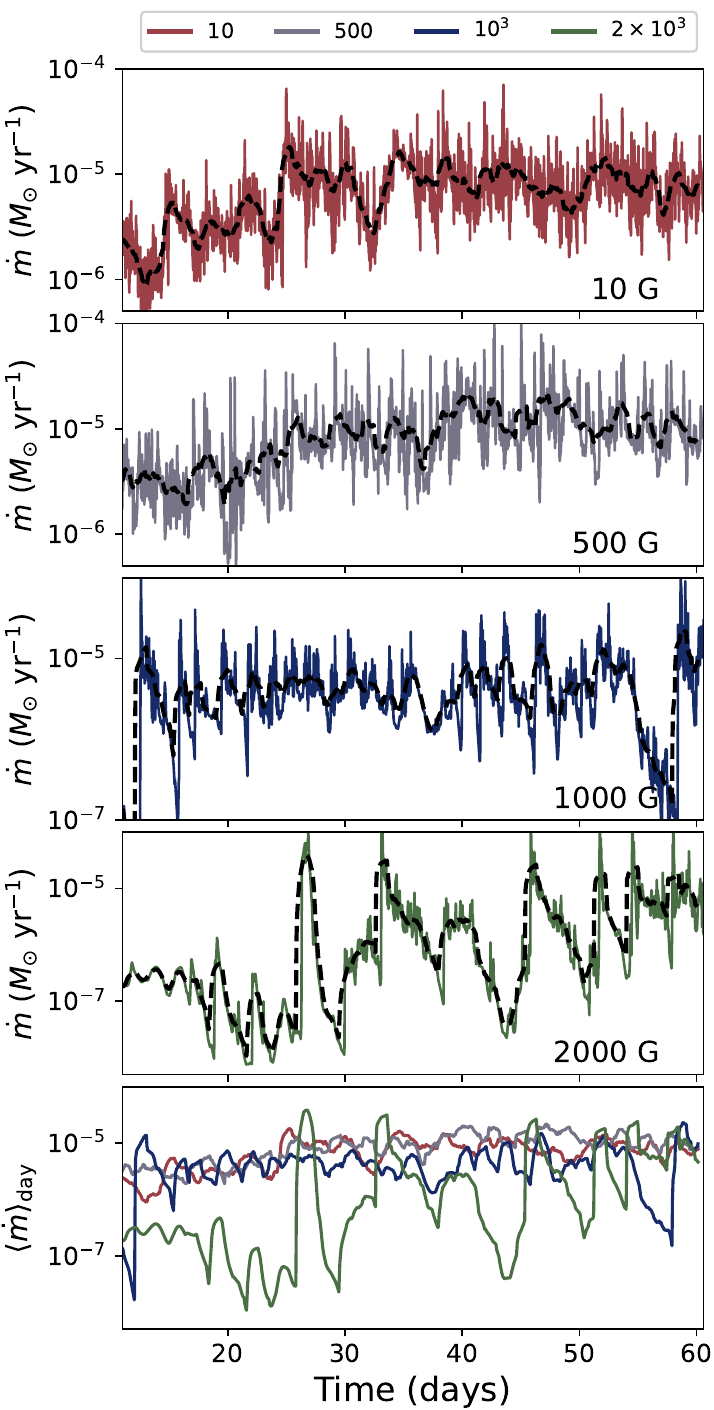}
    \caption{\label{fig:macc} Accretion rate in $M_{\odot}$ yr$^{-1}$ versus time for the four different protostar magnetic fields. Top four panels: Instantaneous accretion rate (solid line) with the rolling average over a day-long window (dashed line). Bottom: 1-day rolling average accretion rates for the different protostellar magnetic fields.}
\end{figure}

The evolution of the accretion rate over time, including any periodic variability and stochasticity, is a potential diagnostic of accretion physics. Figure \ref{fig:macc} shows the total accretion rate, $\dot{m}$, in $M_{\odot}$ yr$^{-1}$, for the different simulations. Figure \ref{fig:macc} shows both the instantaneous accretion rate and a day-long rolling average. The simulations exhibit rapid accretion rate spikes sometimes over an order of magnitude due to the turbulent nature of the inner accretion flow. Since the gas flows in the inner disk are stochastic, the gas hitting the surface can occur in clumps, boosting the accretion rate up substantially between each output. However, as is shown by the black dashed lines in the figure, the 1-day rolling averages show substantially less variability for the 10 G, 500 G, and 1 kG models. There is substantial systematic variability in the 2 kG model, since its accretion is dominated by a bursting mode.

There are clear differences in the accretion rate evolution between the four different cases. Most importantly, the variability of the accretion flow onto the protostar is dependent on the protostar's magnetic field strength. In the lowly magnetized protostar, accretion occurs often via clumps in the turbulent flow. As the protostar's magnetic field is increased, the magnitude of the spikes decreases, as does their occurrence rate. This is due to the protostellar magnetic field launching an outflow and the strong ordered field reducing the turbulence in the accretion region. Therefore, the rapid variability decreases with increasing magnetic field strength. However, for the strongest magnetic field models, 1 kG and 2 kG, a more systematic variability begins occurring. At 2 kG, this systematic variability becomes dominant over the rapidly fluctuating component, demonstrating that the model is in a regime dominated by the protostellar magnetic field. The cause of this variability is discussed further in detail below.

\begin{figure*}
    \centering
    \includegraphics[width=0.2\textwidth]{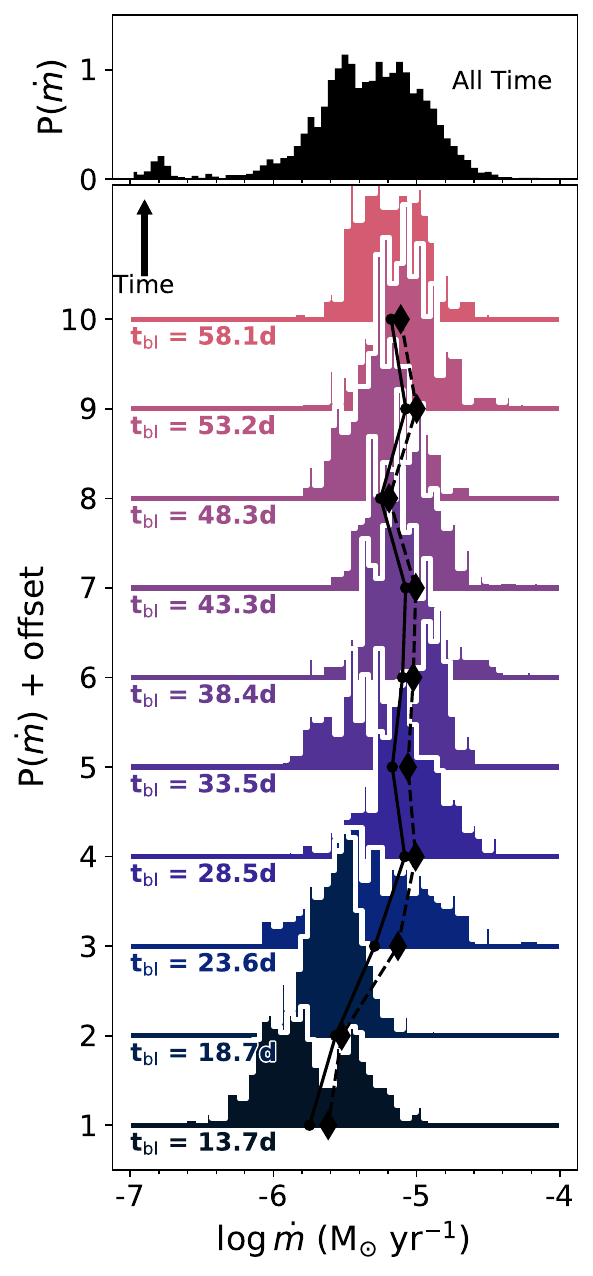}
    \includegraphics[width=0.2\textwidth]{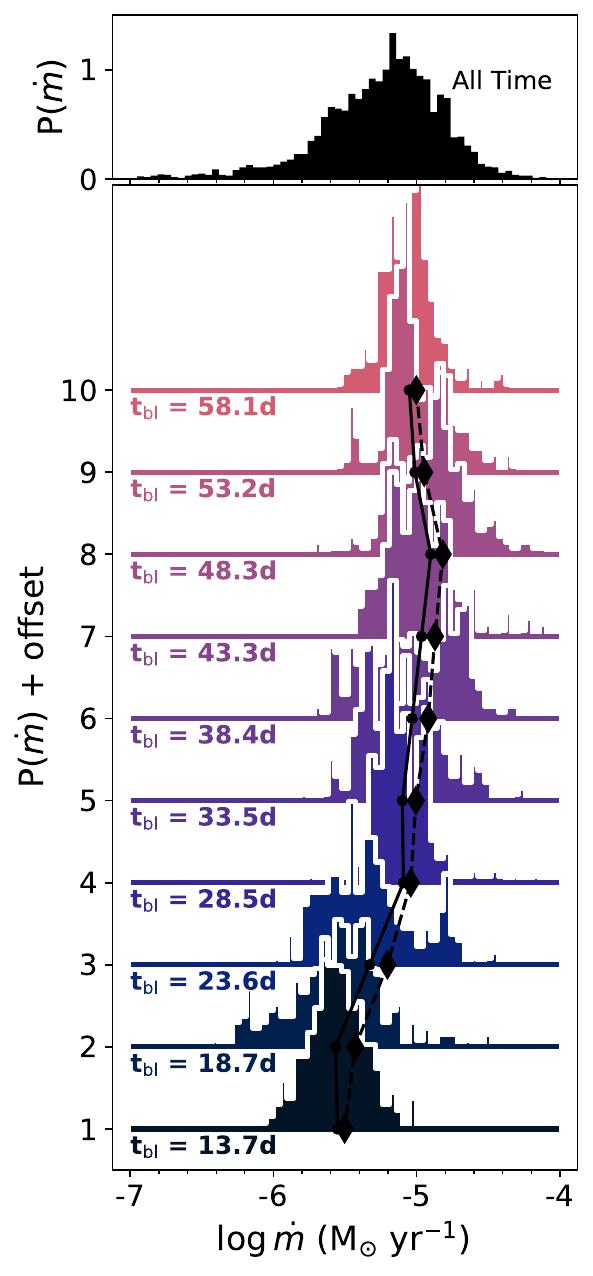}
    \includegraphics[width=0.2\textwidth]{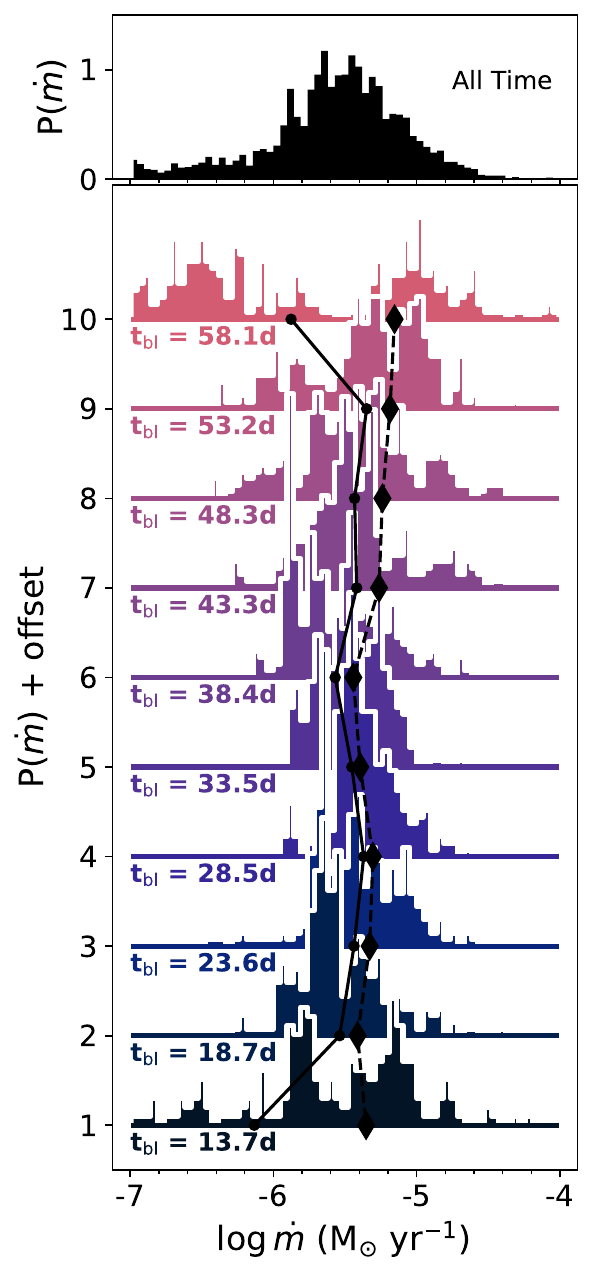}
    \includegraphics[width=0.2\textwidth]{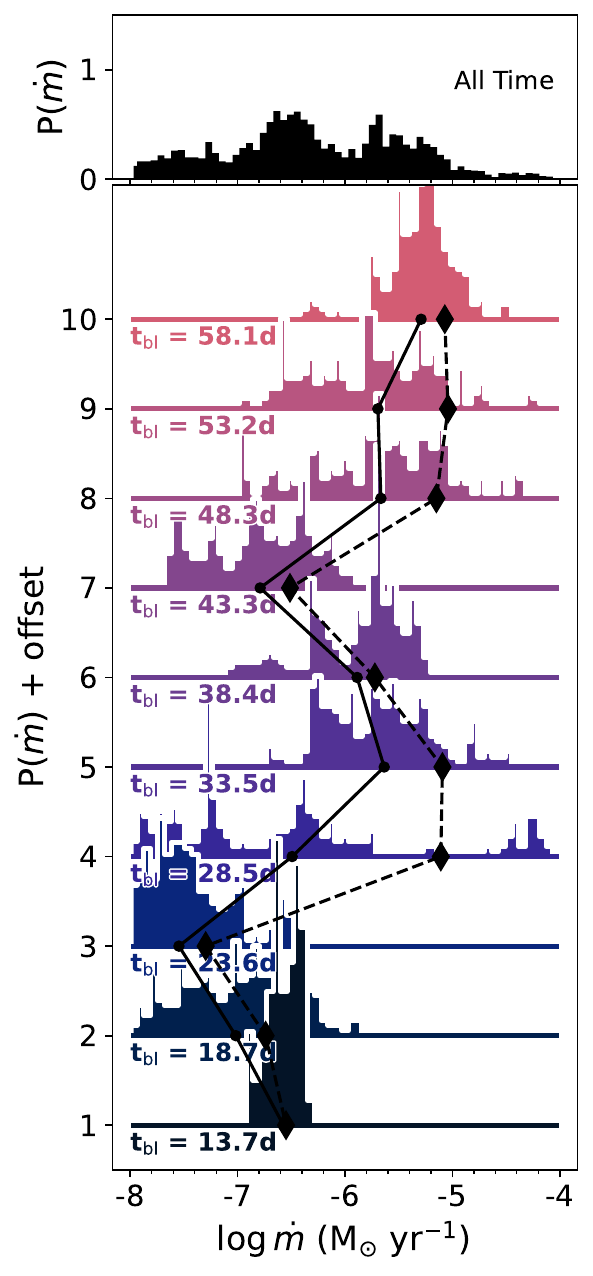}
    \caption{\label{fig:accHists} Distribution of the total accretion rate over $\approx$ 5-day blocks for the 10 G (left) 500 G (middle left) and 1000 Gauss (middle right) and 2000 Gauss (right)protostars. The black points are the logarithmic average of the accretion rate and the diamonds are the time-averaged, e.g., $\langle \dot{m}_* \rangle_{\rm bl} = m_{\rm acc}/T_{\rm window}$, where $m_{\rm acc}$ is the accretion mass during that block. Annotated is the block number (also denoted by color), and the time in the center of the block in days.}
\end{figure*}

Figure \ref{fig:accHists} shows the distribution of accretion rates in 5-day blocks of simulation time for the different magnetic fields. The top sub-panels of the figure show the total accretion rate distribution across the entire time considered for analysis. For the 10 G, 500 G, and 1 kG cases, the distribution broadly follows a log-normal, although the center shifts slightly between these. In the 1 kG model, the distribution flattens at later times, impacted by the substantial drop in the accretion rate resulting from the protostar briefly entering a strong propeller phase. For the 2 kG case, there is a bimodal distribution peaking at higher and lower accretion values, with a tail toward high accretion rates. The lower accretion rate peak corresponds to the stable phase and the higher peak with the tail corresponds to the times when matter dumps from the truncation radius onto the protostellar surface with a subsequent decay in the accretion rate back to the stable phase. 

Figure \ref{fig:accHists} also shows two different solid black points corresponding to the logarithmic average, $\langle \log \dot{m}_* \rangle$ (black circle), and linear time-window average, $\langle \dot{m} \rangle_{\rm bl} = m_{\rm acc}/T_{\rm window}$ (black diamond). The lower three magnetic field strength cases are generally similar indicating that accretion via the stable mode is dominant (e.g., not dominated by major bursts), although the 1 kG case starts to show some differences at later times. For the 2 kG case, there is substantially more mass being accreted during the bursts than the quiescent phase, and as such there is a substantial deviation of up to an order of magnitude between $\langle \log \dot{m}_* \rangle$ and  $\langle \dot{m} \rangle$.

\subsection{Accretion bursts for protostars with kilogauss magnetic fields}\label{sec:burst}
The 2 kG protostar model exhibits clearly defined accretion variability with a quasi-periodic signal defined by strong bursts (periods when the accretion rate grows rapidly). Further, the 1 kG protostar exhibits pseudo-periodic accretion bursts. In both of these models, the protostellar disk is truncated with gas flowing along magnetic flux tubes to the protostellar surface. In this subsection, we present, first, two different limiting regimes to estimate the critical magnetic field to truncate the disk, and then we discuss possible mechanisms to explain the accretion bursts.

\subsubsection{Critical magnetic field strengths for disk truncation}
The critical magnetic field at which the magnetic pressure truncates the disk can be described by balancing the ram pressure of the accreting gas ($\rho v_r^2$) with the magnetic pressure of the protostellar dipole plus the disk magnetic field ($B^2/8\pi$). The radius at which this occurs is called the truncation radius, $R_T$, since the magnetic pressure halts the disk inflow, truncating the disk and producing a distinct separation between the disk and the accretion zone. We consider two different limiting regimes, i) the gas is infalling near free-fall velocities, and ii) the gas is advecting through the disk purely due to viscosity.

For the first limiting regime, we assume the gas is infalling at some fraction, $f_v$, of the free-fall velocity ($v_{\rm ff} = \sqrt{2Gm_*/R_*}$) through some fraction, $f_{\rm acc}$, of a sphere around the protostar. The covering fraction is defined as $f_{\rm acc} = A_{\rm acc}/A_*$ where $A_{\rm acc}$ is the spherical surface area covered by the accretion flow and  $A_* = 4 \pi R_*^2$ is the surface area of the protostar. We denote the magnitude of the total magnetic field as $B = b B_*$, where $b = |\vec{B}_{\rm disk} + \vec{B}_*|/|\vec{B}_*|$, to account for both the protostellar dipole and the disk magnetic field. In the case of anti-parallel fields, $b$ may be less than 1. From this, we can write the critical magnetic field strength needed to make the radius at which the magnetic pressure can compete with the ram pressure outside the protostar radius as
\begin{multline}
    (bB_*)_M  \approx 1 \, {\rm kG} \left ( \frac{f_v}{1.0} \right )^{1/2} \left ( \frac{f_{\rm acc}}{0.1}\right )^{-1/2} \left ( \frac{R_*}{3 \,R_{\odot}} \right )^{-5/4}\\ \times \left ( \frac{\dot{m}_*}{10^{-6}\,{M_{\odot} {\rm yr}^{-1}}}\right )^{1/2} \left ( \frac{m_*}{M_{\odot}}\right )^{1/4}.
\end{multline}

For the second limiting regime, we assume that the accretion onto the protostar is occurring purely via viscous flows through the disk. The geometric area of the accretion flow through the disk can be written as $A \approx f_D R h$, where $f_D$ is the fraction of the disk cross-section through which gas is flowing, $h = c_s/\Omega_K$ is the scale height under equilibrium, $c_s$ is the sound speed and $\Omega_K$ is the Keplerian rotation frequency. The in-fall velocity through the disk, from mass conservation, is $v_r = \frac{\dot{m}_*}{2\pi\Sigma(R) R}$ where $\Sigma \approx \dot{m}/(3\pi\eta_s)$. Using the approximation of a steady-state viscous accretion disk, determining the temperature as a balance of radiative cooling and viscous heating, and an $\alpha$-viscosity given by $\eta_s = \alpha c_s h$, the critical magnetic field strength can be written as
\begin{multline}
    (bB_*)_\alpha \approx 184 \, {\rm Gauss} \left ( \frac{f_V}{1.0} \right )^{1/2} \left ( \frac{f_D}{0.1}\right )^{-1/2} \left ( \frac{\dot{m}_*}{10^{-6}\,{M_{\odot} {\rm yr}^{-1}}}\right )^{9/16} \\ \left ( \frac{R_*}{3 \,R_{\odot}}\right )^{-19/16} \left ( \frac{\alpha}{0.1} \right )^{1/2} \left ( \frac{m_*}{M_{\odot}}\right )^{1/16}.
\end{multline}
As is shown, $(bB_*)_\alpha$  is typically lower than $(bB_*)_M$ since the free fall velocity is greater than the disk accretion velocity. In our simulations, we find that the gas is flowing at sub-free-fall speeds, but faster than the viscous speeds, agreeing with the disk becoming truncated above $B_* > 500$ G. We note that the above equations are approximate since they follow the assumptions that the disk is geometrically thin and the mass and angular momentum transport are in a steady-state equilibrium. However, they provide a rough estimate of this critical field strength.

In the 1 kG and 2 kG simulations, gas builds up at the truncation radius in a stable knot until its mass penetrates the magnetosphere as a filament and funnels onto the protostar, traveling along nearby magnetic fields to the surface. We find that the gas does not hit exactly at the equator and is instead guided by field lines that connect to the surface at low altitudes. 

\subsubsection{Accretion bursting mechanisms in the kilogauss simulations}
We considered two plausible mechanisms to understand the bursting demonstrated in Figure \ref{fig:macc}. For the first, we built a simple analytic prescription in which gas accreting through the disk builds up at the truncation radius until the momentum of the gas can overcome the magnetic pressure, funneling all the gas onto the surface in a burst. We compared the predictions from this model with one of the major bursts from the 2 kG simulation. The density of the gas being built up at the truncation radius is described by by
\begin{equation}
    \rho_{\rm ms} = \frac{\dot{m}_*\times t_b}{f_A \times 2\pi R_{\rm T} \Delta r^2},
\end{equation}
where $f_A$ is the filling fraction of the torus around the protostar (in our simulations, $f_A = 1$ by definition since we do not include the azimuthal dimension), $\Delta r$ is the radial width of the torus, and $t_b$ is the time since the last accretion event. Assuming mass conservation outside of the truncation radius, the radial velocity of the gas flowing against the stable knot of gas is $v_r = f_v v_{\rm ff}$, where $f_v$ is the fraction of gas in-fall speed to the free-fall speed, the time between bursts is
\begin{equation}
 t_b \approx\frac{B^2 f_A (R_{\rm T} \Delta r )^2}{8 G f_V^2 m_* \dot{m}},
\end{equation}
where $B$ is the local magnetic field magnitude. For the burst around day 45 of the simulation, we find that the velocity of the gas in the disk midplane is not free-falling but moving at approximately 10\% free-fall, consistent with the average accretion rate. The width of the torus can be estimated by assuming the gas falling in hits an impenetrable wall and oscillates around that spot in a stable Keplerian orbit, such that width is balanced by the length scale corresponding to a sound wave traveling over a full orbit. The width of the torus then follows
\begin{equation}
    \Delta r \approx c_s \times \frac{2\pi}{\Omega_K(R_T)},
\end{equation}
where $c_s$ is the thermal sound speed and $\Omega_K(R_T)$ is the Keplerian angular speed at the truncation radius. It should be noted that under this assumption, the width of the torus scales similarly to the scale height of the disk at the truncation radius. For a truncation radius, $R_T \approx 0.02$ AU and a gas temperature $\approx 10^4$ Kelvin, the torus width is approximately 0.005 AU, in rough agreement with the empirically found width around the bursts. Using the simulation results around this burst, and $\dot{m}_* = 10^{-6} M_{\odot}$ yr$^{-1}$, the above expression gives $t_b \approx 10$ days, in qualitative agreement with the simulation results.

\begin{figure}
    \centering
    \includegraphics[width=0.95\columnwidth]{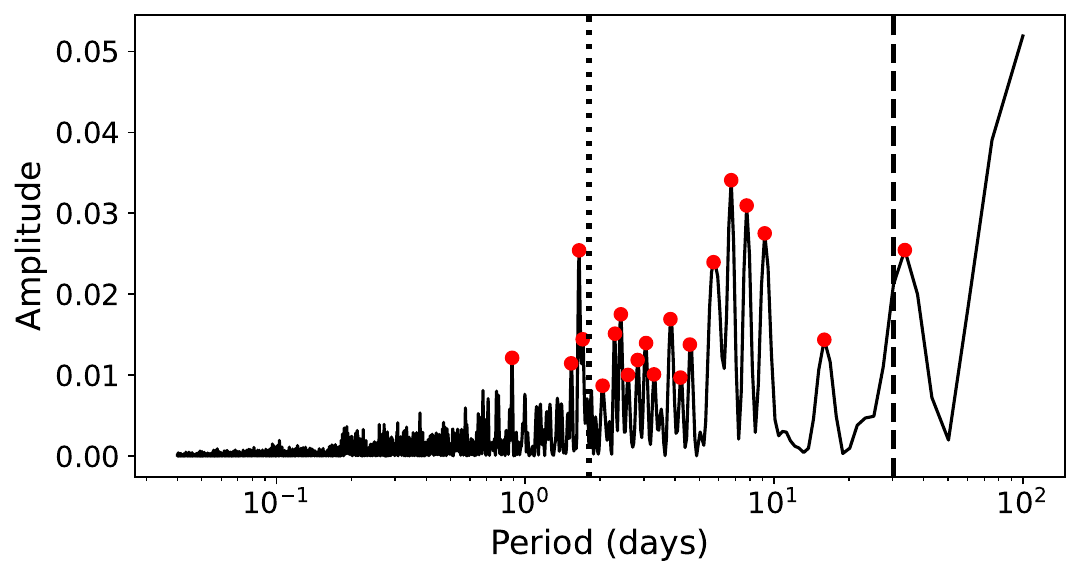}
    \includegraphics[width=0.95\columnwidth]{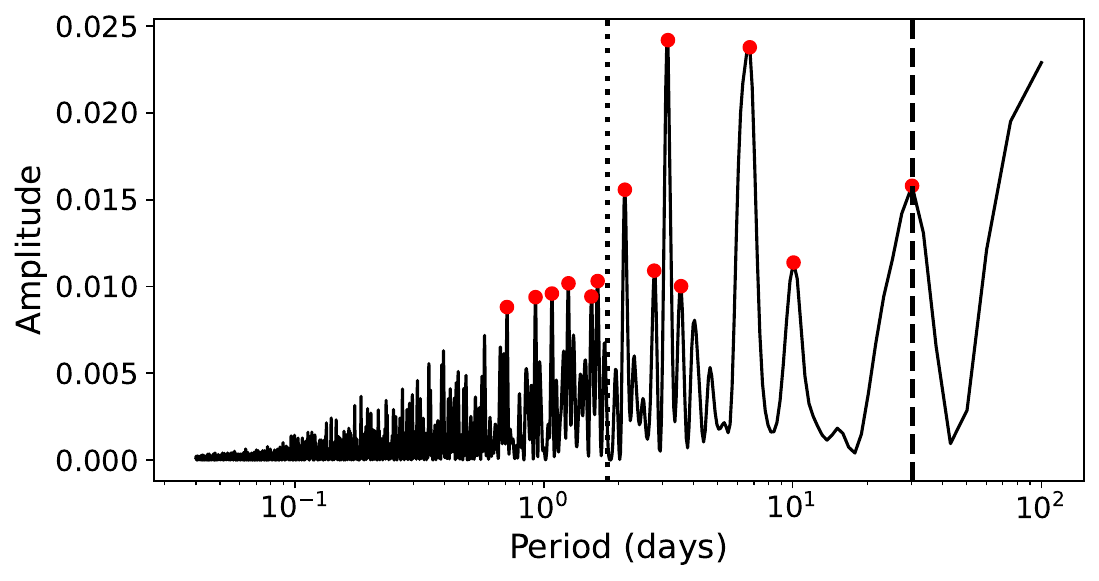}
    \caption{\label{fig:lomb}Lomb-scargle periodogram for the 1 kG (top) and 2 kG (bottom) models. The vertical dotted line shows the stellar rotation period, and the dashed line shows the Nyquist frequency for the duration of the simulation. The red points highlight peaks with a false alarm probability $< 1\%$.}
\end{figure}

However, one would expect that under steady-state background accretion through the disk, the above mechanism would produce a clear signature of periodicity in the accretion rate. Figure \ref{fig:lomb} shows the Lomb-Scargle periodograms for the 1 kG and 2 kG models. We find that there are no clear dominating periodic signals, with power being distributed across a range of frequency peaks. The 2 kG model shows the strongest signal at roughly 3.5 and 7 days, although there is a broad distribution of power at signals from approximately 12 hours to 10 days. 

Due to the lack of clear periodicity, we also consider the magnetic Rayleigh-Taylor instability, also called the ``interchange instability'' \citep{Kruskal1954, Newcomb1961}. This instability has been found numerically to exist in magnetospheric accretion in T-Tauri-like protostars \citep{Kulkarni2008, Blinova2016, Zhu2024}. We utilize the ``fastness'' parameter \citep[e.g.,][]{Ghosh2007, Romanova2018}, defined here as
\begin{equation}
    \omega_s = \frac{\Omega_*}{\Omega(R_T)},
\end{equation}
where $\Omega(R_T) = v_{\phi}/R$ is the rotation rate of the disk at the truncation radius. We consider the general rotation rate rather than the Keplerian rate since, although most of the disk $v_\phi = v_{\rm kep}$, in the inner region there can be departures from Keplerian rotation. We empirically define the truncation radius as the radius at which the total magnetic pressure is balanced by the total ram pressure and thermal pressure \citep{Romanova2018}, 
\begin{equation}
    \frac{B^2}{8\pi} = \rho v^2 + p,
\end{equation}
at $R = R_T$. We use the general expression since the protostar is rotating and thus at any point could be in a propeller regime. Outside of the propeller regime, $\rho v^2$ approaches $\rho v_r^2$ which is often used ram pressure term to balance the magnetic pressure (see above subsection, and review by \citet{Hartmann2016}). \citet{Blinova2016} found from a suite of simulations that the gas at the truncation radius could become unstable to the interchange instability when $\omega_s < 0.6$. While the simulations presented herein include a more massive disk than \citet{Blinova2016}, we consider this definition to roughly estimate when the truncation region may undergo this instability.

\begin{figure}
    \centering
    \includegraphics[width=\columnwidth]{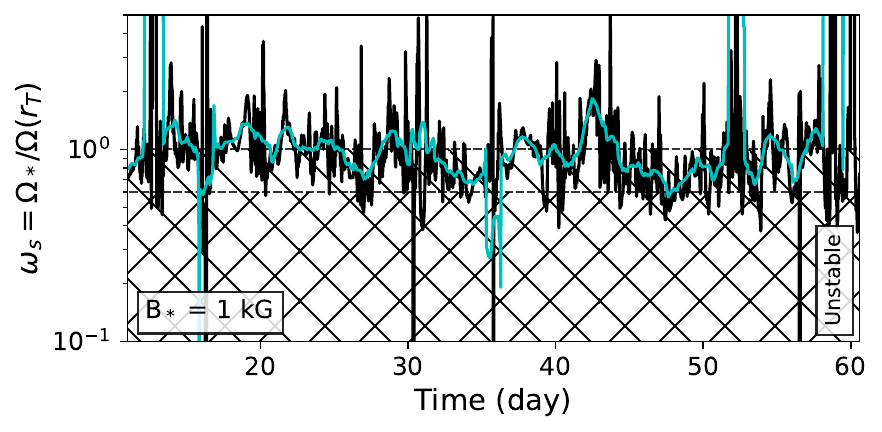}
    \includegraphics[width=\columnwidth]{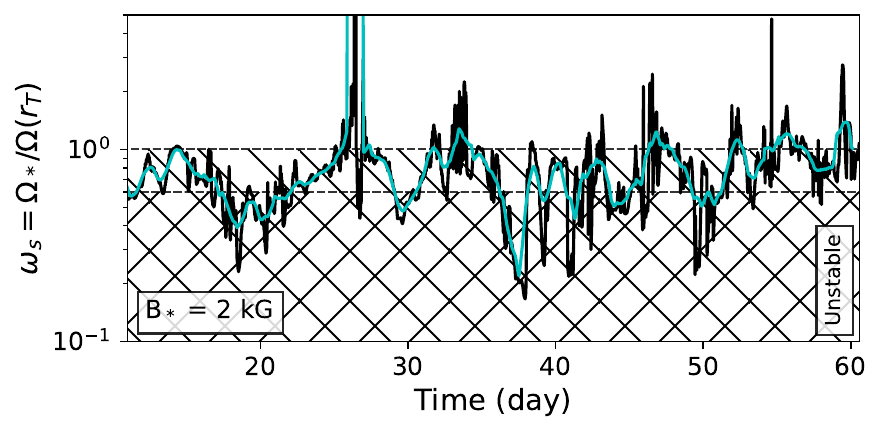}
    \caption{\label{fig:fastness} Fastness parameter, $\omega_s$, as a function of time for the 1 kG (top) and 2 kG (bottom) protostar models. The `\textbackslash` hatched region indicates where $\omega_s < 1$ and the `X` hatched region indicates where $\omega_s < 0.6$, and thus prone to the interchange instability. The cyan line indicates the daily rolling average.}
\end{figure}

Figure \ref{fig:fastness} shows the $\omega_s$ for the 1 kG and 2 kG models. For the 1 kG model, the fastness parameter oscillates significantly between $\omega_s >> 1$ and $\omega_s \approx 1$. While this indicates the protostar is close to the propeller regime, the inner truncation radius does not appear to be strongly in the regime of the instability. Further, the daily rolling average oscillates around $\omega_s \approx 1$. However, it is noteworthy that the period of $\omega_s >> 1$ around day 50 is followed by a significant decrease in the accretion rate for a week. The decrease in the accretion is likely thus due to the protostar being momentarily in the strong propellor regime and the subsequent winds decreasing the accretion rate. 

For the 2 kG case, there is much less rapid stochasticity in $\omega_s$, with a significant amount of time showing $\omega_s \approx 0.6$. Therefore, the pseudo-periodic variability in the accretion rate in this model may be readily explained by the interchange instability. Similarly to the 1 kG case, the only time window in which the rolling average $\omega_s$ greatly exceeds 1 (around day 25) is followed by a significant reduction in the accretion rate. Our results suggest that accretion burst signatures are directly related to the protostellar magnetic field and the star-disk interaction via magnetic coupling. In both cases, the protostar is not entirely within the strong propeller regime, but instead randomly enters it. When this occurs, the magnetosphere expands, suddenly collapsing the gas flow and decreasing the amount of gas in the accretion region, thereby decreasing the accretion rate.

\subsection{Accretion luminosity variability}
\begin{figure}
    \centering
    \includegraphics[width=0.48\textwidth]{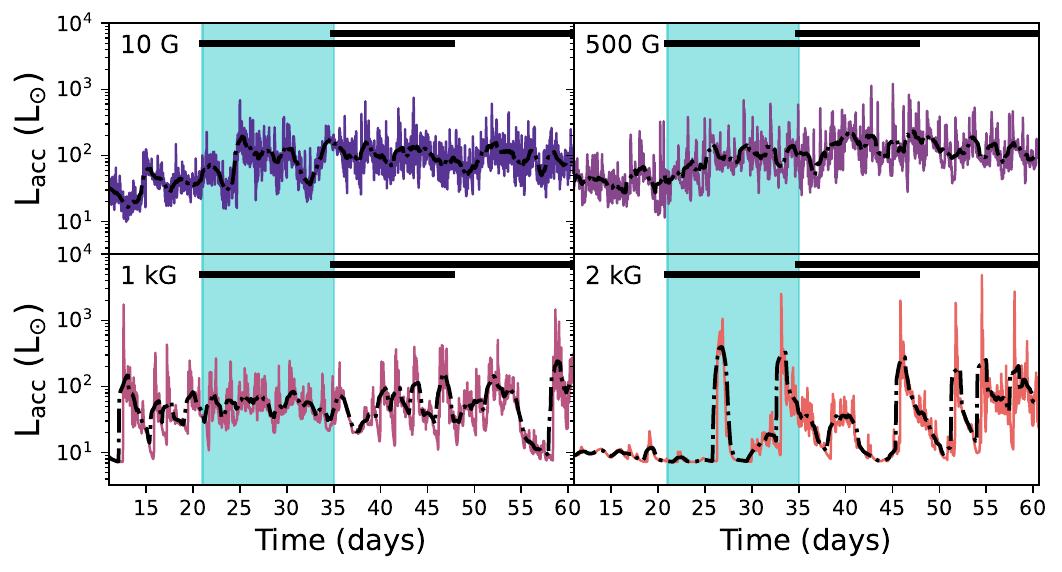}
    \caption{\label{fig:lightCurve}Bolometric accretion luminosity versus time for the different protostellar magnetic field models, annotated on the plots. The cyan band denotes the starting region for the sampled light curves. The black dashed-dot line is the daily rolling average.}
\end{figure}

The luminosity of protostars is observed to vary significantly, with both short and long periods found in Class 0/I protostars \citep[see recent review by][]{Fischer2023}. While the HADES simulations do not trace the monthly or yearly periods, the simulations show significant variability on shorter timescales. The bolometric luminosity here is defined as
\begin{equation}
    L_{\rm bol} = L_* + L_{\rm acc},
\end{equation}
where $L_* = 3.5$ L$_{\odot}$ is the intrinsic luminosity of the protostar, calculated using fits from \citet{Fischer2017} of the stellar tracks from \citet{Siess2000}, and $L_{\rm acc}$ is the accretion luminosity, defined as
\begin{equation}
    L_{\rm acc} = \frac{Gm_* \dot{m}_*}{R_*}.
\end{equation}
We do not consider here individual bands or line emission, nor the reprocessing of the emission (which would reduce the emission as some of the radiation is absorbed by the star). Figure \ref{fig:lightCurve} shows the light curves for each of the four different models.

\begin{figure}
    \centering
    \includegraphics[width=0.48\textwidth]{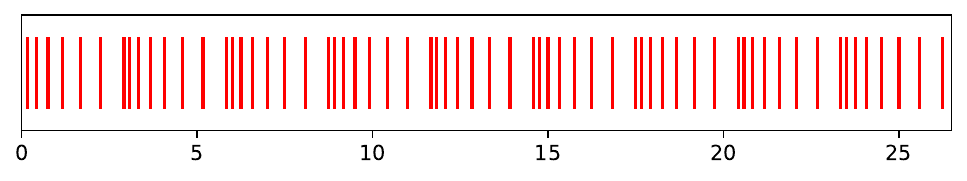}
    \caption{\label{fig:ysovarSamp} Sampling points of the light curve mimicking the YSOVAR survey.}
\end{figure}

The instantaneous light curves exhibit significant differences between the models. In particular, the 10 G and 500 G models have less daily variation in their light curves but have significant variability on rapid timescales of an hour or less. As the magnetic field increases, the hourly variability decreases. For the 1 kG model, the daily average light curve is remarkably flat except for the period after 55 days when the protostar accretion significantly dips. The 2 kG shows a clear systematic variability with much less hourly variability. The emission at early times is dominated by the intrinsic luminosity, and at later times the light curve is dominated by peaks from the accretion bursts. Observationally, unless the protostar is observed multiple times per hour, the 10 G, 500 G, and 1 kG models would exhibit relatively flat light curves, except for the 1 kG model entering the strong propeller regime. The clear variability in the 2 kG exhibits a nearly exponential decrease which may be a clear signature of a highly magnetized protostar accreting via instability. 

Observations of sources in Serpens South by YSOVAR \citep{Wolk2018} exhibit a range of light curves, including some consistent with those here. In particular, they find line curves with rapid but small-scale variability (e.g., SSTYSV J183000.56-020211.4, SSTYSV J183005.33-020111.8, SSTYSV J183005.81-020205.8) and others with larger broader variability (e.g., SSTYSV J183001.26-020148.3). Their object SSTYSV J183006.13-020108.0 shows a significant dip in luminosity, which they calculate has a period of approximately 35 days. However, this matches their sampling period, so it may be that the dip is not periodic variability, but instead from catching a protostar dipping due to entering the strong propeller regime and leaving it.

\begin{figure}
    \centering
    \includegraphics[width=0.5\textwidth]{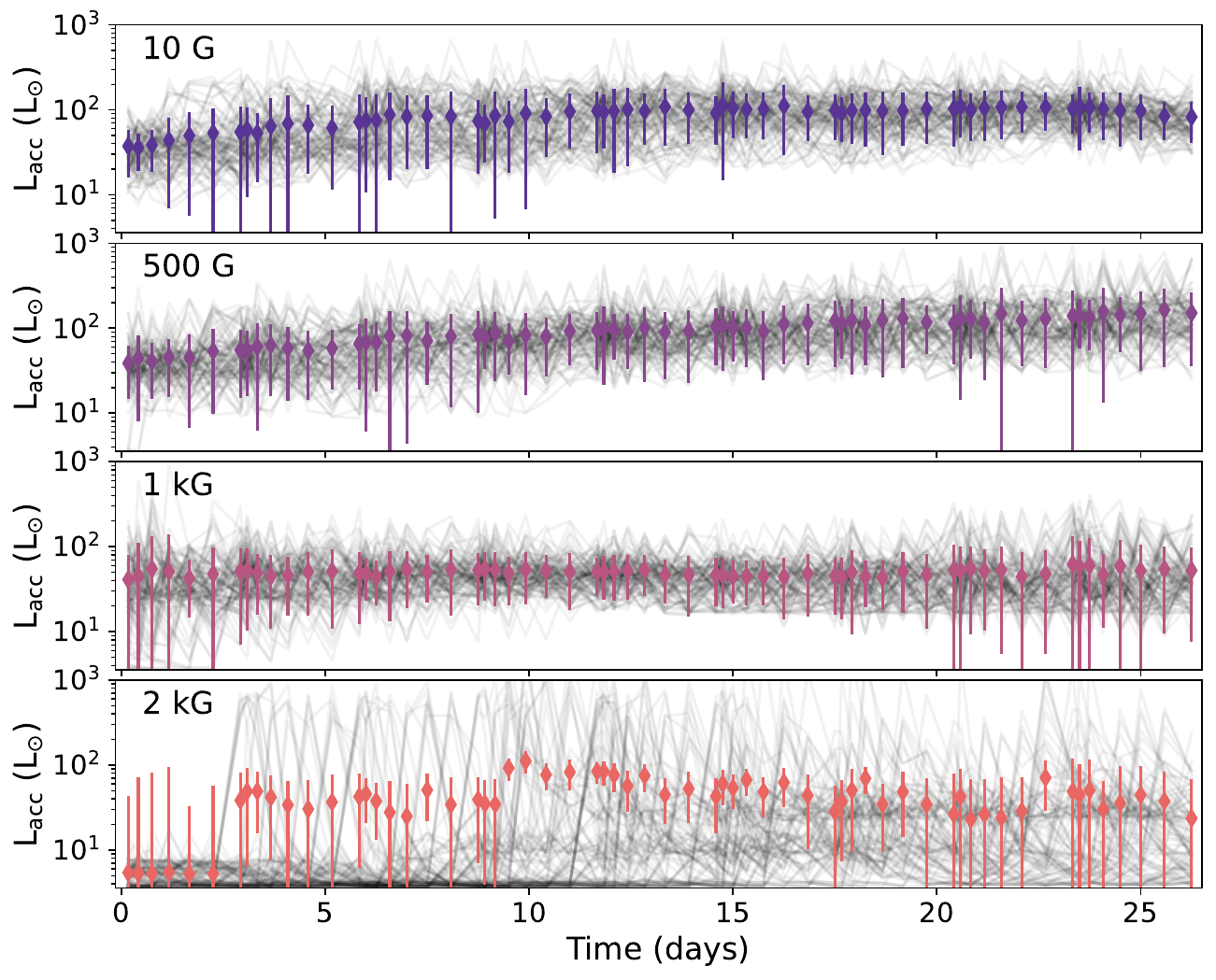}
    \caption{\label{fig:lightsamp}Synthetic light curve samples (black points) with the average and standard deviations given by the colored points.}
\end{figure}

\begin{figure}
    \centering
    \includegraphics[width=0.5\textwidth]{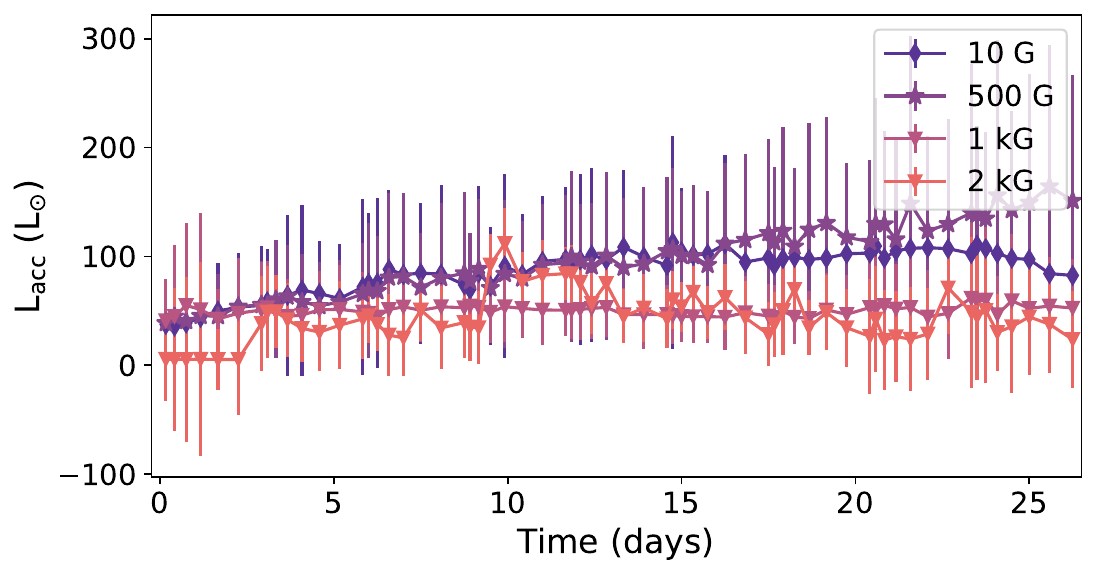}
    \caption{\label{fig:avgLightCurve}Average light curves from the samples in Fig. \ref{fig:lightsamp} with the standard deviation.}
\end{figure}

Observational campaigns to study protostellar variability are restricted in their sampling rate. One of the most robust surveys is YSOVAR \citep{Rebull2014, Wolk2015, Wolk2018}, which utilized a logarithmic-spaced sampling strategy to trace both shorter and longer periods. We mimic the sampling strategy of YSOVAR, as is shown in Figure \ref{fig:ysovarSamp}, to produce synthetic light curve samples of our models. To produce a light curve for an ``average protostar'' similar to our models, we compute a synthetic ``population average'' YSOVAR light curve. We choose random starting points for the ``observation'' of the light curve, uniformly distributed in the blue bands shown in Figure \ref{fig:lightCurve}. Starting from this initial time, we generate a light curve sampled following the Figure \ref{fig:ysovarSamp} sampling strategy to generate a population of light curves. Figure \ref{fig:lightsamp} shows these samples along with the logarithmic average and standard deviations of this population. The figure shows there is substantial scatter in the individual samples, with some sampled light curves showing very high luminosities. However, the average behavior is substantially milder. Figure \ref{fig:avgLightCurve} highlights the logarithmically averaged light curves and the standard deviations at each sample point. The variabilities, rather than orders of magnitude, are only factors of a few. Defining the magnitude variation as $\Delta M = \log_{2.5} \sigma(\langle L_{\rm acc, samp} \rangle)$, we find $\Delta M \approx$ 0.3, 0.39, 0.12 and 0.78 for the 10 G, 500 G, 1 kG, and 2 kG models, respectively. These are in agreement with the variability measurements in \citet{Wolk2018}. Much of the accretion variability is similar in this regard, however, for the 1 kG case it is smaller due to the more ordered fields which are not strong enough yet to induce the large variability seen in the 2 kG. In general, we find the 500 G model can become the brightest due to the enhanced accretion caused by the outflow cavity focusing the accretion flow, while the strongly magnetized sources are the dimmest.

Unfortunately, the results of the sampling mean that even with the incredible cadence of YSOVAR, constraining the hourly variability will be difficult. The order of magnitude variability in Figure \ref{fig:lightCurve} is substantially reduced, on average, at the cadence sampling of YSOVAR. The YSOVAR sampling strategy is highly useful if one wishes to trace both shorter and long-term evolution with minimal samples. Instead, a campaign using a rapid ($< 1$ hour) linearly spaced sampling strategy over several nights would be necessary to constrain any rapid variability.

\subsection{Accretion filling fraction}
\begin{figure}
    \centering
    \includegraphics[width=0.5\textwidth]{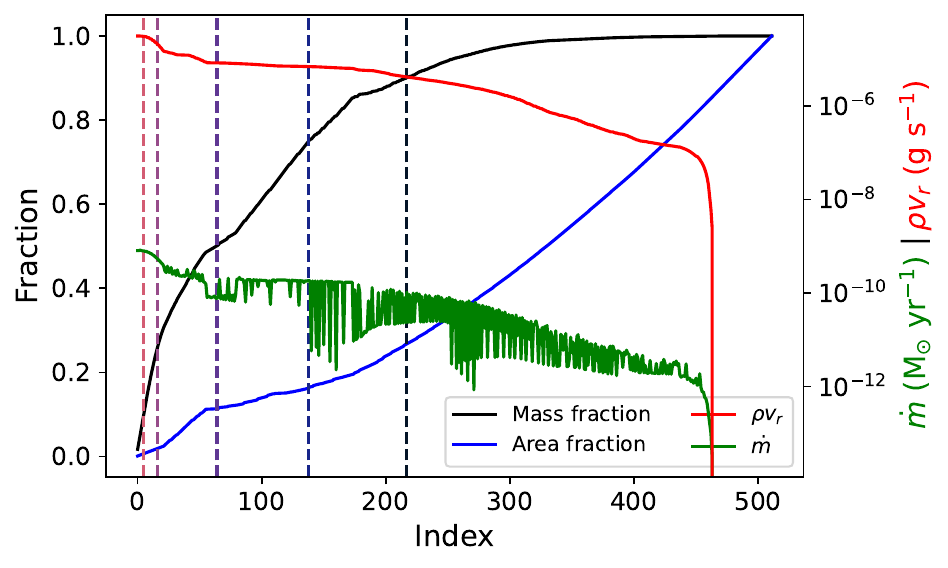}
    \caption{\label{fig:mflux_i} Example of how the filling fraction is computed for a single snapshot. The figure shows the sorted mass flux at the surface (red), accretion rate (green), cumulative mass accretion (black), and cumulative area (blue) versus the sorted cell index. The horizontal lines show the index at which the top 10\%, 25\%, 50\%, 75\%, and 90\% of the mass is accreted.}
\end{figure}

\begin{figure}
    \centering
    \includegraphics[width=0.5\textwidth]{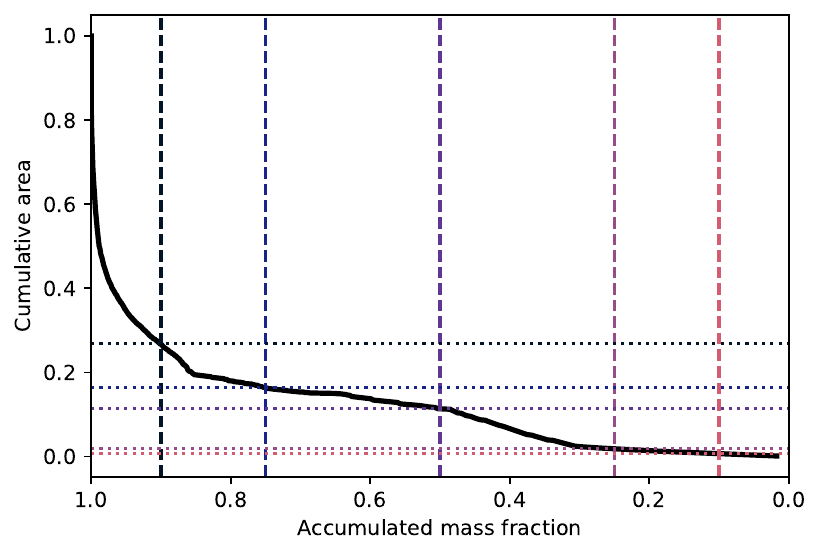}
    \caption{\label{fig:mflux_area} Accumulated mass fraction percentage versus the cumulative area of the protostar surface for the snapshot in Figure \ref{fig:mflux_i}. The horizontal lines show the index at which the top 10\%, 25\%, 50\%, 75\%, and 90\% of the mass is accreted.}
\end{figure}

The filling fraction of the accretion onto the protostar is of key importance for models of star formation and protostar evolution. We computed the filling fraction of accretion for the top 10, 25, 50, 75, and 90\% of the accreted mass as a function of time. To compute the filling fraction, for each snapshot, we sort the cells along the protostellar surface according to the mass flux, from highest flux to lowest. Marching down from the cells with the most mass being accreted to the least enables a better constraint on the effective area where a given percentage of the mass is being accreted. Using the flux and the cell surface area, we calculate the cumulative mass accretion fraction. Then, for the different top mass-accretion percentages we compute the total area of the cells marching down in mass flux which encompasses that percentage of the total accretion. Figure \ref{fig:mflux_i} shows an example of a single snapshot from the 2 kG model, plotting the sorted mass flux versus the index, along with the cumulative mass accretion, area, and the cell's accretion rate. For this snapshot, Figure \ref{fig:mflux_area} shows the accumulated mass fraction (marching up from the lowest) versus the area fraction with lines denoting the 10, 25, 50, 75, and 90\% for this snapshot example. The figure shows that the vast majority of the mass is accreted over a minority of the surface. 

\begin{figure}
    \centering
    \includegraphics[width=0.5\textwidth]{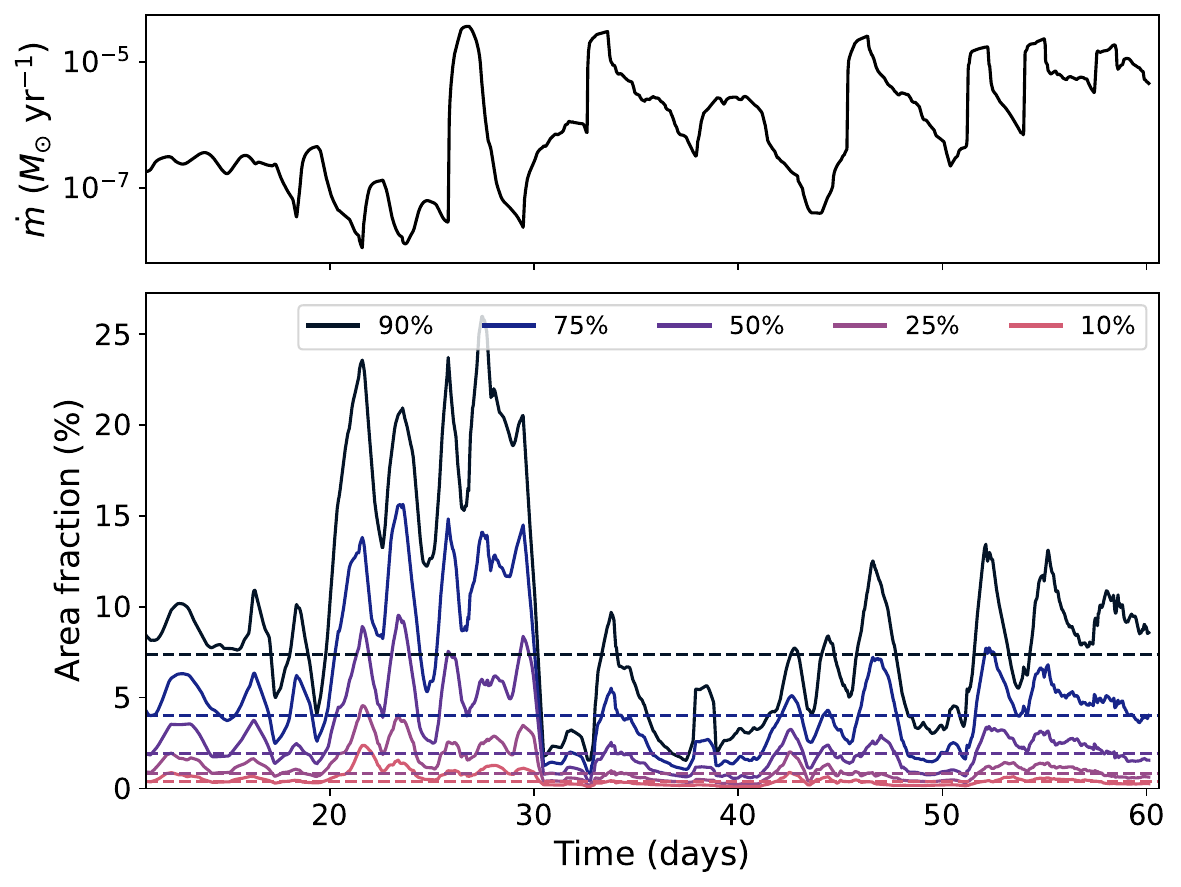}
    \caption{\label{fig:fillFrac_B2000} Top: The accretion rate versus time for the 2 kG protostar, using the daily-rolling average. Bottom: The daily rolling average filling fraction of the accretion flow for the different top mass accretion percentages. }
\end{figure}

Figure \ref{fig:fillFrac_B2000} shows the filling fraction of accretion versus time for the 2 kG protostar, plotted using the daily rolling average for ease of visualization and interpretation. Table \ref{tab:fillingFrac} shows a summary of the accretion-rate-weighted average filling fraction over the simulation period for the four different models. We find that the top 25\% of the mass is accreted in about 1\% of the surface area. The majority of the mass, at 50\%, is accreted over 2-4\% of the surface. Finally, of most importance, 90\% of the accreted mass is from mass flux encompassing between 7 - 20\% of the surface. In all models, the last 10\% of the mass is in general accreted more diffusely over the surface (see e.g., Figure \ref{fig:mflux_area}, where the last 10\% of the mass is accreted over 70\% of the area). In general, we find that these filling fractions are roughly equal with time, however, they can increase by factors of 2-3 during accretion bursts, as is seen in Figure \ref{fig:fillFrac_B2000}.  Depending on the protostellar magnetic field assumed, a filling fraction of protostar accretion between 10-20\% is reasonable for protostars accreting under these modes. 

\begin{table}
    \caption{Filling fraction (percent) of the accretion flow onto the protostar according to the top X\% of the mass accretion.}
    \label{tab:fillingFrac}
    \centering
    \begin{tabular}{r|r r r r r}
        \hline \hline
        B$_*$ & 10\% & 25\% & 50\% & 75\% & 90\% \\
        \hline
        10 G   & 0.4 & 1.0 & 2.5 & 6.6 & 15.9 \\
        500 G  & 0.6 & 1.5 & 4.2 & 10.2 & 19.7 \\
        1000 G & 0.5 & 1.2 & 3.1 & 6.8 & 12.1 \\
        2000 G & 0.4 & 0.8 & 1.9 & 4.0 & 7.4 \\
        \hline
    \end{tabular}
\end{table}

\section{Discussion}\label{sec:disc}
After analyzing these simulations, we find that there are three distinct regimes of the accretion: i) boundary layer (10 G, 500 G), ii) turbulent magnetospheric (1 kG), and iii) magnetospheric accompanied by the interchange instability (2 kG).   

\begin{figure}
    \centering
    \includegraphics[width=0.5\textwidth]{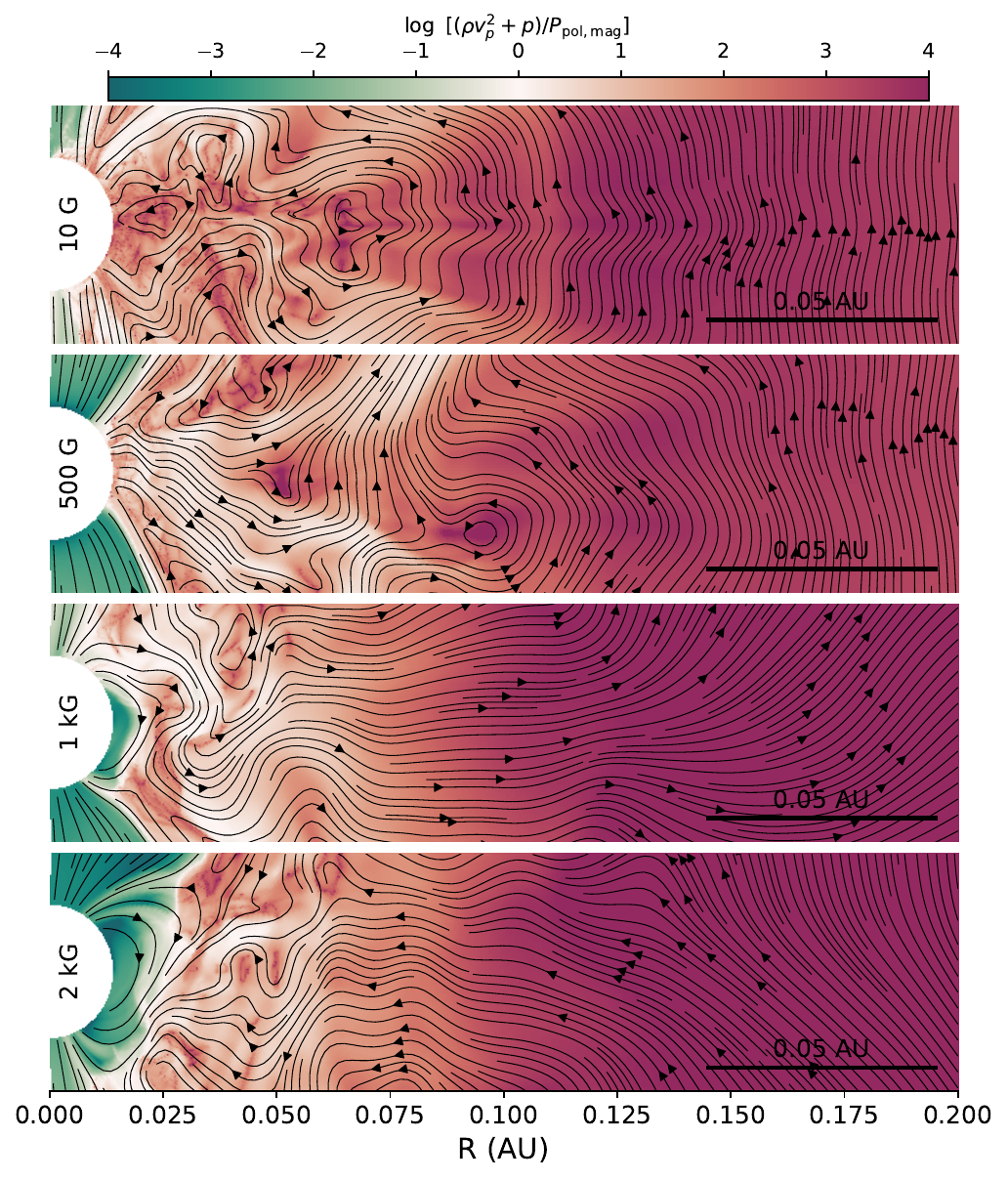}
    \caption{\label{fig:magSum} Ratio of the poloidal ram pressure plus thermal pressure to the poloidal magnetic pressure within the inner 0.2 AU of the disk. The black arrows indicate the direction of the poloidal component of the magnetic field. The ratio is averaged over 1 day starting from day 50. The panels are in ascending order of protostar magnetic field, annotated on the top, of 10 G, 500 G, 1 kG, and 2 kG from top to bottom.}
\end{figure}

Figure \ref{fig:magSum} visualizes the importance of the magnetic field via the ratio of the poloidal pressures, $(\rho v_p^2 + p)/[(B_r^2 + B_\theta^2)/2]$, in the inner region of the disk. The figure shows the average ratio from days 50-51 of the simulation and the lines denote the direction of the poloidal magnetic field. The 10 G model shows the gas fully impacts the protostar surface with only a marginally magnetically dominated pole. The 500 G model immediately demonstrates the impact of the magnetically dominated cavity creating a pressure that inhibits gas flow to the poles, focusing the accretion to the equator. The kilogauss-strength models show clearly magnetically truncated disks. In all cases, toward outer radii (0.2 AU), the magnetic fields transition to vertical. In the inner disk, there is a substantial difference between the models, especially due to the impact of the Hall effect for the cases with strong magnetospheres. Inside the disk, there is a strong correlation between the morphology of the field lines, the presence of magnetic islands, and the ratio of the pressures.

\citet{Takasao2018} simulated an accretion disk around an unmagnetized central object, with the T-Tauri scaling providing a nominal accretion rate of $\dot{m}_* \approx 10^{-8} M_{\odot}$ yr$^{-1}$, a factor of 100 less than in this study. They find ordered boundary layer accretion with an accretion component due to high-latitude funnel wall flows. This is in broad agreement with our 10 G simulation, where we find that the protostellar disk directly impacts the protostar surface, with two high latitude, nearly constant, inflows due to failed disk winds cycling back down to the disk surface and sliding along the surface. However, the accretion variability we see is significantly higher than theirs (see their Fig. 19), which may be due to the difference in the disk surface density and the overall accretion rate magnitude difference. Mass ejections from the protostar surface can disrupt the accretion momentarily, leading to dips in the averaged accretion rate. \citet{Takasao2018}, similarly to these simulations, find that the midplane component, especially at later times, becomes complemented by higher latitude flows. However, the high latitude flows in the results presented herein occur at $\theta > 30^\circ$, rather than 15$^\circ$ they find. While the initial density distributions between the simulations differ, both simulations use hour-glass morphologies for the disk magnetic field.

Our simulations also tell us why the boundary layer accretion in the 500 G protostar is ordered and enhanced. The protostellar magnetic field induces a strongly magnetized, ordered bipolar outflow. This magnetized cavity produces a pressure pushing the inner accretion flow down, constraining and focusing it toward the protostar's equator. While the presence of the protostellar magnetic field does not truncate the disk and create well-defined accretion columns, the resulting outflow orders the accretion flow away from the protostellar poles, funneling it toward the equation, boosting the accretion rate compared to the 10 G protostar. As a protostellar magnetic field is increased, the accretion flow will become more ordered, transitioning from a turbulent boundary layer mode to a more ordered boundary layer mode, eventually to the magnetospheric mode at kilogauss scales with bursts.

\section{Conclusions}\label{sec:conclude}
We have presented the initial suite of high-resolution simulations of accretion disks around embedded, actively accreting protostars. We modeled protostars with four different protostellar magnetic fields: 10 G, 500 G, 1 kG, and 2 kG. These simulations resolve the inner accretion disk with a maximal resolution of $\approx$ 10$^{-4}$ AU. Our simulations show that

\begin{enumerate}
    \item The protostellar magnetic field plays a crucial role in determining the underlying accretion mechanism. We find that for solar-mass protostars accreting at rates exceeding $10^{-7}$ M$_{\odot}$ yr$^{-1}$, protostars must generate a magnetic field exceeding 1 kG to truncate the disk and produce magnetospheric accretion. For weaker fields, we find that boundary layer accretion dominates.
    \item For poorly magnetized protostars (10 G), the accretion falls across the entire surface of the protostar. The gas impacts both at the equator and via two high-latitude streamers that are the result of failed winds falling back in.
    \item For marginally magnetized protostars (500 G), in which the disk is not truncated, gas accretes onto the protostar via boundary layer accretion. However, the presence of a magnetically driven bipolar outflow focuses the accretion flow toward the protostar's equator, enhancing and regulating the accretion rate.
    \item For kilogauss strength fields, we find that there is substantial systematic variability, with the 2 kG model exhibiting strong bursts of accretion. The most likely explanation for the accretion bursts in the 2 kG is the interchange instability occurring at the truncation radius. 
    \item Our 1 kG and 2 kG simulations have one-sided outflows due to the symmetry of the protostellar magnetic dipole being broken by the disk magnetic field.
    \item In our simulations, the protostars accrete the vast majority of the mass across 10-20\% of their surface area. While there is some variance in the filling fraction between the protostellar magnetic fields, for protostar accretion modeling and sub-grid models we would recommend a filling fraction of 10-20\% for protostars at these stages of evolution.
\end{enumerate}

We investigated the variability in the accretion luminosity for the four different models. The instantaneous light curve found in our simulation varies by up to over an order of magnitude depending on the magnetic field. For the 10 G, 500 G, and 1 kG models, the variability generally occurs at hourly timescales, while the $B_* = 2$ kG model exhibits significant variability of over an order of magnitude over days due to accretion bursts. The bolometric accretion luminosity computed this way does not account for radiation transfer effects. Observations of variability do not sample uniformly for sub-hourly timescales continuously but instead employ a range of strategies such as the logarithmic sampling of YSOVAR. Therefore, we produced averaged synthetic light curves using a cadence to mimic YSOVAR \citep{Rebull2014}. We find that there is far less variability in these average light curves, with bolometric magnitude variances less than one, which is roughly consistent with observations \citep{Wolk2018}.

These simulations motivate further observations of magnetic fields toward Class I protostars. The recent survey from \citet{Flores2024} measured the magnetic fields for a range of Class I sources using high-resolution K-band spectroscopy. Their sample exhibits magnetic fields ranging from 500 G to nearly 4 kG, with the majority of the sample having fields between 1 - 2 kG. While the survey is not sensitive to unmagnetized protostars, a key takeaway is that the magnetic field does exhibit a broad spread, and as such our simulation results across a range of protostellar magnetic fields can aid in interpreting accretion observations of these earlier protostars. Future work will investigate synthetic hydrogen recombination line emission, a broader parameter space for the protostar rotation, accretion rate, disk magnetic field morphologies, and the physics of outflow launching.

\begin{acknowledgements}
BALG is supported by the Chalmers Initiative on Cosmic Origins as a Cosmic Origins Postdoctoral Fellow. JCT acknowledges support from ERC Advanced Grant 788829 (MSTAR). RK acknowledges financial support via the Heisenberg Research Grant funded by the Deutsche Forschungsgemeinschaft (DFG, German Research Foundation) under grant no.~KU 2849/9, project no.~445783058. The computations were enabled by resources provided by the National Academic Infrastructure for Supercomputing in Sweden (NAISS) at Dardel/PDC partially funded by the Swedish Research Council through grant agreement no. 2022-06725 through the allocations SNIC 2022/5-654 and NAISS 2023/23-341 and resources provided by Chalmers e-Commons at Chalmers.

This work made use of the following software packages: \texttt{matplotlib} \citep{Hunter:2007}, \texttt{numpy} \citep{numpy} and \texttt{python} \citep{python}. Parts of the results in this work make use of the colormaps in the \texttt{CMasher} \citep{2020JOSS....5.2004V, CMasher_10677366} and \texttt{cmocean} \citep{cmocean}. Software citation information aggregated using \texttt{The Software Citation Station} \citep{software-citation-station-paper, software-citation-station-zenodo}.
\end{acknowledgements}

\bibliographystyle{aa}
\bibliography{lib}

\end{document}